\documentclass[12pt]{article}
\usepackage{amsmath}
\usepackage{amssymb}
\usepackage{graphicx}
\usepackage{enumerate}
\usepackage[longnamesfirst]{natbib}
\usepackage{xcolor}
\usepackage{url} 
\usepackage{xr}

\usepackage{fancyhdr}
\newcommand{\blind}{1}

\newdimen\defpicwidth
\newdimen\defepswidth
\def\SKparam#1#2{}

\long\def\@makecaption#1#2{%
  \vskip\abovecaptionskip
  \sbox\@tempboxa{#1. #2}%
  \ifdim \wd\@tempboxa >\hsize
    {\centerline{\renewcommand{\baselinestretch}{0.8}%
\small\normalsize\parbox{0.8\textwidth}{#1. #2}}}\par
  \else
    \global \@minipagefalse
    \hbox to\hsize{\hfil\box\@tempboxa\hfil}%
  \fi
  \vskip\belowcaptionskip}

\newcommand{\CO}{{\mathcal{O}}}
\def\Co{{\scriptstyle{\mathcal{O}}}}

{\begin{longtable}{|p{1mm}p{3mm}p{0.8\textwidth}p{1mm}|}
\hline\hline
&&&\\[-4mm]
&& \hskip-20pt \centerline{\bf \summaryname}&\\[1mm]
\hline\hline\endfirsthead
\hline\hline
&&&\\[-4mm]
&& \hskip-20pt \centerline{\bf \summarycontname }&\\[1mm]
\hline\hline\endhead
\hline\hline\endfoot
}
{\end{longtable}}

\providecommand{\pdf}[1]{}

\newcommand{\dirxplore}[1]{../#1}
\newcommand{\dircodes}[1]{\wwwebook/codes/#1}

\def\wwwebook{http://www.quantlet.de}

\def\definitionname{DEFINITION}

\def\proofname{PROOF}

\def\remarkname{REMARK}
\def\summarycontname{Summary (continued)}
\def\summaryname{Summary}

\def\bookname{}

\ifx\sfb\undefined
  \newcommand{\sfb}{}
  \fi

  \def\bookname{sfs}
\SKparam{BEGIN}{xpl ps pdf}
\SKparam{INPUT}{titlesv}
\SKparam{END}{}
\SKparam{BEGIN}{html java}
\SKparam{INPUT}{matitcs}
\SKparam{END}{}
\SKparam{LIB}{sfs}
\def\wwwebook{http://www.quantlet.com/mdstat}

\newcommand{\E}{\mathop{\mbox{\sf E}}}     

\SKparam{BEGIN}{ps pdf}
\SKparam{INPUT}{dedic}
\SKparam{INPUT}{pref}
\SKparam{END}{}

\renewcommand{\P}{\mathrm{P}}            

\newenvironment{proof}{\noindent{\bf \proofname.}}
{\quad ${\Box}$ \newline}

\def\proofname{PROOF}

\def\remarkname{REMARK}

\newtheorem{theorem}{THEOREM}[section]

\newtheorem{lemma}{LEMMA}[section]

\newtheorem{remark}{\remarkname}[section]
\newtheorem{definition}{\definitionname}[section]

\newtheorem{assumption}{Assumption}[section]
\newtheorem{prop}{Proposition}[section]

\makeatletter
\newcommand*{\addFileDependency}[1]{
  \typeout{(#1)}
  \@addtofilelist{#1}
  \IfFileExists{#1}{}{\typeout{No file #1.}}
}
\makeatother

\begin{document}

\def\spacingset#1{\renewcommand{\baselinestretch}%
{#1}\small\normalsize} \spacingset{1}


\if1\blind
{
  \title{\bf A projection based approach for interactive fixed effects panel data models}
  \author{Georg Keilbar \hspace{.2cm}\\
    Humboldt-Universit\"at zu Berlin, Chair of Statistics, Germany\\ email: georg.keilbar@hu-berlin.de \\
    Juan M. Rodriguez-Poo \\
    University of Cantabria, Department of Economics, Spain\\ email: juan.rodriguez@unican.es \\
    Alexandra Sober\'on \\
    University of Cantabria, Department of Economics, Spain\\ email: alexandra.soberon@unican.es \\
    Weining Wang \\
    University of Bristol, School of Economics, UK\\ email: weining.wang@bristol.ac.uk}
  \maketitle
} \fi

\if0\blind
{
  \bigskip
  \bigskip
  \bigskip
  \begin{center}
    {\LARGE\bf A projection based approach for interactive fixed effects panel data models}
\end{center}
  \medskip
} \fi

\begin{abstract}
This paper introduces a straightforward sieve-based approach for estimating and conducting inference on regression parameters in panel data models with interactive fixed effects. The method's key assumption is that factor loadings can be decomposed into an unknown smooth function of individual characteristics plus an idiosyncratic error term. Our estimator offers advantages over existing approaches by taking a simple partial least squares form, eliminating the need for iterative procedures or preliminary factor estimation.
In deriving the asymptotic properties, we discover that the limiting distribution exhibits a discontinuity that depends on how well our basis functions explain the factor loadings, as measured by the variance of the error factor loadings. This finding reveals that conventional ``plug-in'' methods using the estimated asymptotic covariance can produce excessively conservative coverage probabilities. We demonstrate that uniformly valid non-conservative inference can be achieved through the cross-sectional bootstrap method. Monte Carlo simulations confirm the estimator's strong performance in terms of mean squared error and good coverage results for the bootstrap procedure. We demonstrate the practical relevance of our methodology by analyzing growth rate determinants across OECD countries.

\end{abstract}

\noindent%
\vfill

\newpage
\spacingset{1.9} 

\section{Introduction}\label{sec:intro}
In this paper, we consider the consistent estimation of the following panel data model with interactive fixed effects, often referred to as a factor structure model. These models have gained significant attention over the past decade due to their ability to control latent variables and account for cross-sectional dependence. The model we consider takes the following form,
\begin{align}\label{eq1}
    \begin{array}{ll}
    y_{it}&=\mathbf{X}_{it}^{\top}\boldsymbol{\beta}+v_{it},
    \\
    v_{it}&=\boldsymbol \lambda_i^{\top}\mathbf{f}_t+u_{it},
    \end{array}
    \qquad i=1,\ldots,N;\quad t=1,\ldots,T,
\end{align}
where $y_{it}$ is the response variable of individual $i$ at time $t$, $\mathbf{X}_{it}$ is a $Q$-dimensional vector of covariates, and $\boldsymbol{\boldsymbol{\beta}}$ is the $Q$-dimensional vector of parameters to be estimated. The error term $v_{it}$ has two components: (i) a latent factor structure $\boldsymbol{\boldsymbol \lambda}_i^{\top}\mathbf{f}_t$, where $\boldsymbol{\boldsymbol \lambda}_i=(\lambda_{i1},\ldots,\lambda_{iK})^{\top}$ represents individual specific factor loadings and $\mathbf{f}_t=(f_{t1},\ldots,f_{tK})^{\top}$ denotes common unobserved factors; (ii) an idiosyncratic error term, $u_{it}$, which is assumed to be zero mean and independent of the covariates and the factor structure. The number of factors $K$ is finite and does not depend on the size of the cross-section $N$ or the time dimension $T$.

A critical challenge of model (\ref{eq1}) is the potential correlation between covariates $\mathbf{X}_{it}$ and latent factors, $\mathbf{f}_t$, or the individual factor loadings, $\boldsymbol{\boldsymbol \lambda}_i$. This correlation introduces cross-sectional dependence (CSD) in the structural model, and standard estimation techniques for panel data models can lead to inconsistent estimates and spurious inferences of the parameters of interest because they do not account for the latent structure embedded in the error term $v_{it}$. In this context, this paper aims to provide an innovative estimation procedure to deal with this endogeneity issue and obtain consistent estimates of the slope coefficient vector $\boldsymbol{\boldsymbol{\beta}}$.

Addressing this issue is crucial in many areas of economics, where some of the regressors are decision variables that are influenced by unobserved individual heterogeneities. For instance, in macroeconomics, cross-country growth studies often seek to understand the determinants of economic growth, $y_{it}$, using observable variables, $\mathbf{X}_{it}$, such as physical capital investment, population growth, and trade openness. However, global shocks, such as financial crises or fluctuations in world oil prices affect all countries, introduce common latent factors that affect all countries simultaneously through trade and financial links and must be taken into account to obtain valid estimates \citep{Chudik_Mohaddes_Pesaran_Raissi2017, lu2016shrinkage}. Another relevant example is found in financial econometrics, where excess returns $y_{it}$ for asset $i$ at time $t$ are often modeled using observable factors such as Fama-French factors (e.g., small market capitalization and book-to-market ratios), dividend yields and payout ratios, among others. However, asset returns are also influenced by unobserved global financial conditions, such as the state of the global credit market, which induce a strong CSD across assets \citep{Bernanke-Booivin-Eliasz2005, Fan-Ke-Liao2021}.

To address this source of bias in large panels (where both $N$ and $T$ are large), two main strands have emerged in the literature: one focuses on controlling for common factors, $\mathbf{f}_t$, while the other focuses on modeling factor loadings, $\boldsymbol{\boldsymbol \lambda}_i$. The literature on controlling for common factors is extensive. A widely used methodology is the Common Correlated Effects (CCE) estimator introduced by \citet{Pesaran2006}, which approximates unobserved factors through linear combinations of cross-sectional averages of both dependent and explanatory variables. While this approach offers computational simplicity and broad applicability, its reliance on asymptotic properties may lead to biased estimates in finite samples (see \citet{Westerlund-Urbain2013} for a deeper discussion). An alternative methodology, developed by \citet{Bai2009} and further examined by \citet{Moon-Weidner2015, Moon-Weidner2017}, adopts a fundamentally different approach. Their method employs the principal component (PC) approach to directly estimate the unobserved factors and later compute consistent estimates for $\boldsymbol{\beta}$ solving a non-convex optimization problem. Although this PC-based approach provides an elegant solution, it presents practical challenges due to its computational intensity and sensitivity to the number of unobserved factors. After these original papers, a large body of literature emerged by extending these estimation procedures to more general settings. See \citet{Sarafidis_Wansbeek2012}, \citet{chudik_pesaran2015}, or \citet{Bai-Wang2016} for excellent surveys, and \citet{Westerlund-Urbain2015} for a comparison analysis between the CCE and PC estimation procedures. 

The second strand of the literature focuses on modeling the factor loadings $\boldsymbol{\boldsymbol \lambda}_i$ as a function of observed time-invariant covariates $\mathbf{Z}_{i}$. \citet{Connor-Linton2007}, \citet{Connor-Hagmann-Linton2012}, \citet{Zhang_Zhou_Wang2021}, and \citet{Cheng_Dong_Gao_Linton2024} propose modeling the factor loadings as $\lambda_{ik}=g_k(\mathbf{Z}_{i})$ for some unknown function $g_k(\cdot)$ and $k=1,\ldots,K$. However, this modeling imposes a restrictive assumption that the entire variation in $\boldsymbol{\boldsymbol \lambda}_i$ must be explained by $\mathbf{Z}_{i}$, which increases the risk of model misspecification. To allow for partial flexibility and mitigate the risk of misspecification, \citet{Fan-Liao-Wang2016} propose a more flexible framework that decomposes the factor loadings into a systematic component and an idiosyncratic error:
\begin{align}\label{eq3}
    \boldsymbol{\boldsymbol \lambda}_i=\mathbf g(\mathbf{Z}_i)+\boldsymbol{\boldsymbol \gamma}_{i}, \quad i=1,\ldots,N,
\end{align}
where $\mathbf{Z}_i$ is a $D$-dimensional vector of additional covariates of individual characteristics, $\mathbf g(\mathbf{Z}_i)=(g_1(\mathbf{Z}_i),\ldots,g_K(\mathbf{Z}_{i}))^{\top}$ is a $K$-dimensional vector of unknown functions, and $\boldsymbol{\boldsymbol \gamma}_i=(\gamma_{i1},\ldots, \gamma_{iK})^{\top}$ is a $K$-dimensional vector of errors that reflects the part of $\boldsymbol{\boldsymbol \lambda}_i$ that cannot be explained by $\mathbf{Z}_{i}$. Throughout the paper, it is assumed that $\{\boldsymbol{\boldsymbol \gamma}_i\}_{i\leq N}$ is zero mean and independent of $\{\mathbf{Z}_{i}\}_{i\leq N}$.

Building on this literature, this paper introduces a novel estimation methodology, very easy to implement from the empirical point of view, that leads to consistent estimators of $\boldsymbol{\beta}$ in (\ref{eq1}) when $\mathbf{X}_{it}$ are correlated with $\boldsymbol{\boldsymbol \lambda}_i$ and/or $\mathbf{f}_{t}$. More precisely, inspired by \citet{Fan-Liao-Wang2016} we propose to introduce the relationship in (\ref{eq3}) into a panel data framework, as Eq. (\ref{eq1}). Hence, plugging (\ref{eq3}) into (\ref{eq1}) and rearranging terms, we get the following regression model 
\begin{align}\label{eq4}
    y_{it}=\mathbf{X}_{it}^{\top}\boldsymbol{\boldsymbol{\beta}}+\mathbf g(\mathbf{Z}_{i})^{\top}\mathbf{f}_t+\boldsymbol{\boldsymbol \gamma}_i^{\top}\mathbf{f}_t+u_{it}, \qquad i=1,\ldots,N;\quad t=1,\ldots,T.
\end{align}

This new framework addresses the bias caused by the term $\mathbf{g}(\mathbf{Z}_{i})^{\top}\mathbf{f}_t + \boldsymbol{\boldsymbol \gamma}_i^{\top}\mathbf{f}_t$, which introduces endogeneity in the ordinary least squares (OLS) estimator. Hence, the innovation of this paper lies in projecting the data onto the subspace generated by sieve basis functions of the covariates $\mathbf{Z}_{i}$. This orthogonal projection intends to ``project away'' the unobserved factor loadings to eliminate the bias asymptotically and obtain consistent and asymptotically normal estimators of the $\boldsymbol{\boldsymbol{\beta}}$'s in (\ref{eq4}) without the need for computationally intensive procedures.

Our model setup is closely related to the one in \citet{Zhang_Zhou_Wang2021}, however, we want to point out crucial distinctions. First and foremost, the main issue of interest in the above paper is efficiency and they propose a GLS-type estimator that under broadly general conditions is oracle efficient. It is important to note that their asymptotic results require the consistency of the pooled OLS estimator in a first step which is not the case in our model setup. Second, \citet{Zhang_Zhou_Wang2021} assume that the factor loadings are fully explained by $\mathbf{Z}_i$, i.e., $\boldsymbol{\gamma}_i = 0$. Unfortunately, in the presence of error factor loadings, i.e., $\boldsymbol{\gamma}_i\ne 0$, the statistical properties of standard estimators for $\boldsymbol{\boldsymbol{\beta}}$ remain unclear. Therefore, it is of interest to derive a new estimator to obtain consistency and asymptotic rates.

In this framework, the estimation procedure proposed in this paper offers several advantages. First, it is based on a simple OLS framework, which avoids the complexities of iterative procedures such as the PC method with unknown convergence properties that may be computationally intensive. Second, it does not require prior knowledge of the number of common factors and does not require knowledge or assumptions about them, making it robust to various specifications. Further, the underlying limiting distribution is centered at zero. Therefore, the proposed estimation techniques circumvent the omitted variable bias problem as in \citet{Bai2009}. Finally, the proposed estimator reaches the semiparametric efficiency bound under certain conditions. 

We want to highlight that the asymptotic results of the proposed estimator hold irrespectively of the variance of the error factor loadings being zero, close to zero, or much larger than zero. However, there exists a discontinuity in the limiting distribution when this variance is close to zero. Then, the usual ``plug-in'' approaches would lead to valid but overly conservative inference. Similarly, ignoring the idiosyncratic part leads to invalid inference in the case of persistent variance. To achieve uniformly valid but non-conservative inference, we resort to the cross-sectional bootstrap originally proposed by \citet{kapetanios2008bootstrap}. In this way, by stacking the time observations we are able to mimic the asymptotic distribution and conduct uniformly valid inference even in the presence of this type of discontinuities as shown by \citet{liao2018uniform} and \citet{FernndezVal2022DynamicHD}. The issue of uniformity is an important topic for modeling panel data. \citet{lu2023uniform} consider a model with two-dimensional heterogeneity of varying degrees in the slope parameters and are interested in uniformly valid inference. \citet{kock2016oracle} and \citet{kock2019uniform} study uniform inference in high-dimensional panel regression contexts. \citet{menzel2021bootstrap} showed that uniform non-conservative inference is impossible under general dependence in more than one dimension. The novelty of our bootstrap procedure is that we resample cross-sectional units after projecting the data, i.e., partialing-out the modeled part of the factor loadings.

The rest of the paper is organized as follows. In Section 2, we derive our projection-based interactive fixed effects estimator. Section 3 states our assumptions and studies the asymptotic properties of the proposed estimators. In Section 4 we validate the theoretical results in a simulation study. In Section 5 we apply our method to the identification of the determinants of economic growth. \citet{lu2016shrinkage} argued that the GDP growth rates per capita might not only be determined by observed factors but might also be influenced by latent factors or shocks. Our projection-based interactive fixed effects estimator is well suited for such a setting. All proofs of the asymptotic results and further Monte Carlo results are relegated to a Supplementary Material document.

\section{Estimation Procedure}\label{sec2}
To nonparametrically estimate the unknown function $g_k(\cdot)$ without curse of dimensionality, it will be assumed that for each $k$, where $k=1,\ldots,K$, $g_k(\cdot)$ is an additive function of the form
\begin{align}\label{add}
    g_k\left(\mathbf{Z}_{i}\right) = \sum^D_{d=1}g_{kd}\left(Z_{i,d}\right),\quad i=1,\ldots,N,\quad k=1,\ldots,K.
\end{align}

For each $k$ and $d$, the additive component $g_{kd}(\cdot)$ can be approximated by the sieve method. We define $\left\{\phi_1(Z_{i,d}), \ldots, \phi_{J_N}(Z_{i,d})\right\}$ as a set of basis functions (i.e., splines, Fourier series, wavelets), which spans a dense linear space of the functional space for $g_{kd}(\cdot)$. Then,
\begin{align}\label{eq6}
    g_{kd}(Z_{i,d}) = \sum_{j=1}^{J_N} b_{j,kd}\phi_{j}(Z_{i,d}) +R_{kd}(Z_{i,d}), \quad k = 1, \ldots, K, \quad d = 1,\ldots, D,
\end{align}
where, for $j=1,\ldots,J_N$, $\phi_{j}(\cdot)$'s are the sieve basis functions, $b_{j,kd}$'s are the sieve coefficients of the $d$th additive component of $g_k(\mathbf{Z}_{i})$ corresponding to the $k$th factor loading, $R_{kd}(\cdot)$ is a ``remainder function'' that represents the approximation error, and $J_N$ denotes the number of sieve terms which grows slowly as $N\rightarrow\infty$. 

As it is well-known in the literature,
under some regularity condition of the functional class, the approximation functions $\phi_{j}(\cdot)$ have the property that, as $J_N$ grows, there is a linear combination of $\phi_{j}(\cdot)$ that can approximate $g_k(\cdot)$ arbitrarily well in the sense that the approximation error can be made arbitrarily small. Therefore, for a given $d=1,\ldots,D$, the basic assumption for sieve approximation is that $\sup_z|R_{kd}(z)|\rightarrow0$, as $J_N\rightarrow\infty$. In practice, an optimal choice for the smoothing parameter $J_N$ can be based on cross-validation.

For the sake of simplicity, we take the same basis functions in (\ref{eq6}) and, for each $k\leq K$, $d \le D$ and $i\leq N$, let us define
\begin{align*}
    \mathbf b_k^{\top}&=(b_{1,k1},\ldots,b_{J_N,k1}, \ldots, b_{1,kD}, \ldots, b_{J_N,kD})\in\mathbb{R}^{J_ND},\\
    \phi(\mathbf{Z}_{i})^{\top}&=(\phi_1(Z_{i,1}),\ldots,\phi_{J_N}(Z_{i,1}),\ldots,\phi_1(Z_{i,D}),\ldots,\phi_{J_N}(Z_{i,D}))\in\mathbb{R}^{J_ND},
\end{align*}
so the above equation can be rewritten as
\begin{align}\label{eq7}
    g_k(\mathbf{Z}_{i})=\phi(\mathbf{Z}_{i})^{\top}\mathbf b_{k}+\sum^D_{d=1}R_{kd}(\mathbf{Z}_{i,d}).
\end{align}

Let $\mathbf{Z} = \left(\mathbf{Z}^{\top}_{1},\ldots,\mathbf{Z}^{\top}_{N}\right)$ be an $N\times D$ matrix whose $i$th element is a $D$-dimensional vector of random variables as $ \mathbf{Z}_{i}=(Z_{i1},\ldots,Z_{iD})^{\top}$ and denote by $\mathbf{\mathcal{Z}}$ its support. Let also $\mathbf{G}(\mathbf{Z})$ be an $N\times K$ matrix of unknown functions, $g_k(\mathbf{Z}_{i})$, $\Phi(\mathbf{Z})=(\phi(\mathbf{Z}_{1}),\ldots,\phi(\mathbf{Z}_{N}))^{\top}$ be an $N\times J_ND$ matrix of basis functions, $\mathbf B=(\mathbf b_1,\ldots,\mathbf b_K)$ be a $J_ND\times K$ matrix of sieve coefficients, and $\mathbf R(\mathbf{Z})$ be an $N\times K$ matrix with the $(i,k)$th element $\sum^D_{d=1}R_{kd}(\mathbf{Z}_{i,d})$. By considering (\ref{eq7}) in matrix form, we obtain
\begin{align}\label{eq8}
    \mathbf{G}(\mathbf{Z})=\Phi(\mathbf{Z})\mathbf B+\mathbf R(\mathbf{Z}),
\end{align}
and substituting (\ref{eq8}) into the matrix form of (\ref{eq4}) leads to
\begin{align}\label{eq9}
   \mathbf y_t=\mathbf{X}_t\boldsymbol{\beta}+\Phi(\mathbf{Z})\mathbf B\mathbf{f}_t+\mathbf R(\mathbf{Z})\mathbf{f}_t+\mathbf v_t,\qquad t=1,\ldots,T,
\end{align}
where the residual term consists of two parts: the sieve approximation error, $\mathbf R(\mathbf{Z})\mathbf{f}_t$, and the error term, $\mathbf v_t$.
$\mathbf{v}_t$ is an $N\times 1$ vector such as $\mathbf{v}_t=\boldsymbol \Gamma \mathbf{f}_t+ \mathbf u_t$, where $\boldsymbol\Gamma=(\boldsymbol{\gamma}_1,\ldots,\boldsymbol{\gamma}_N)^{\top}$ is an $N\times K$ matrix of unknown loading coefficients and $\mathbf u_t = \left(u_{1t},\ldots,u_{Nt}\right)^{\top}$ is an $N\times 1$ vector of idiosyncratic errors. Finally, $\mathbf{X}_t$ is an $N \times Q$ matrix of covariates.

To obtain consistent estimators of $\boldsymbol{\beta}$ in (\ref{eq9}) we propose a transformation that removes $\Phi(\mathbf{Z})\mathbf B\mathbf{f}_t$ and accounts for the error term, $\boldsymbol \Gamma \mathbf{f}_t$. A natural choice to remove $\Phi(\mathbf{Z})\mathbf B\mathbf{f}_t$ in (\ref{eq9}) is to define the following projection matrix
\begin{align}\label{eq10}
 \mathbf  P_{\Phi}(\mathbf{Z}) \stackrel{\operatorname{def}}{=} \Phi(\mathbf{Z})\left[\Phi(\mathbf{Z})^{\top}\Phi(\mathbf{Z})\right]^{-1}\Phi(\mathbf{Z})^{\top}.
\end{align}
Premultiplying both sides of (\ref{eq9}) by $\mathbf{P}_{\Phi}(\mathbf{Z})$ and assuming that $(NT)^{-1}\sum_{t=1}^T \mathbf{X}_t^{\top} \left[\mathbf{I}_N- \mathbf P_{\Phi}(\mathbf{Z})\right]\mathbf{X}_t$ is non-singular, the following estimator for $\boldsymbol{\boldsymbol{\boldsymbol{\beta}}}$ is obtained,
\begin{align}\label{eq12}
    \boldsymbol{\boldsymbol{\widehat{\boldsymbol{\beta}}}} = \left\{\frac{1}{NT}\sum_{t=1}^T \mathbf{X}_t^{\top} \left[\mathbf{I}_N- \mathbf P_{\Phi}(\mathbf{Z})\right]\mathbf{X}_t\right\}^{-1}\frac{1}{NT}\sum_{t=1}^T\mathbf{X}_t^{\top}\left[\mathbf{I}_N- \mathbf P_{\Phi}(\mathbf{Z})\right] \mathbf y_t.
\end{align}

\section{Asymptotic Properties}\label{sec3}
In this section, we analyze the main asymptotic properties of the estimator. Firstly, we introduce some notation, definitions, and assumptions that will be necessary to derive the main results of this paper. Later, we present the main large sample properties of these estimators. All proofs of these results are relegated to a Supplementary Material document. 

\subsection{Notation}
Let $n = NT$. For two positive number sequences $(a_n)$ and $(b_n)$, we say $a_n=\CO(b_n)$ or $a_n\lesssim b_n$ (resp. $a_n\asymp b_n$) if there exists $C>0$ such that $a_n/b_n\le C$ (resp. $1/C\le a_n/b_n\le C$) for all large $n$, and say $a_n=\Co(b_n)$ if $a_n/b_n\rightarrow0$ as $n\rightarrow\infty$. We set $(X_n)$ and $(Y_n)$ to be two sequences of random variables. Write $X_n=\CO_{p}(Y_n)$ if for $\forall \epsilon>0$, there exists $C>0$ such that $\P(|X_n/Y_n|\leq C)>1-\epsilon$ for all large $n$, and say $X_n=\Co_{p}(Y_n)$ if $X_n/Y_n\rightarrow 0$ in probability as $n\rightarrow\infty$. We use $\text{plim}$ to denote the probability limit. Further, for a real matrix $\mathbf A$, let $\|\mathbf A\|_F={\rm tr}^{1/2}(\mathbf A^{\top}\mathbf A)$ and $\|\mathbf A\|_2=\lambda_{\max}^{1/2}(\mathbf A^{\top}\mathbf A)$ denote its Frobenius and spectral norms, respectively. Let $\lambda_{\min}(\cdot)$ and $\lambda_{\max}(\cdot)$ denote the minimum and maximum eigenvalues of a square matrix. For a vector $\textbf{v}$, let $\|\textbf{v}\|$ denote its Euclidean norm.

\subsection{Definitions and Assumptions}
\begin{definition}\label{definition}
A function $h(\cdot)$ is said to belong to the class of additive functions 
$\mathcal{G}$, if :
$h(\cdot) = \sum^D_{d=1}h_d(\cdot)$ and $h_d(\cdot)$ belongs to the H\"older class of functions 
\[
    \left\{h_d:|h^{(r)}_d(s)-h^{(r)}_d(t)|\leq L|s-t|^{\zeta}\right\}
\]
for some $L>0$, and for all $s$ and $t$ in the domain of $h_d(\cdot)$, where $r$ stands for the $r$-th derivative of the real-valued function $h_d(\cdot)$ and $0<\zeta\leq 1$.
\end{definition}

For any scalar or vector function $\varphi(z)$, we use the notation $\Pi_{\mathcal{G}}[\varphi(z)]$ to denote the projection of $\varphi(z)$ onto the class of functions $\mathcal{G}$. That is, $\Pi_{\mathcal{G}}[\varphi(z)]$ is an element that belongs to $\mathcal{G}$ and is the closest function to $\varphi(z)$ among all the functions in $\mathcal{G}$. More specifically, we have
\begin{eqnarray}\label{eq17}
\nonumber & & \E\{[\varphi(z)-{\Pi}_{\mathcal{G}}(\varphi(z))][\varphi(z)-
{\Pi}_{\mathcal{G}}(\varphi(z))]^{\top}\}\\
=& & \inf_{h\in\mathcal{G}}\E\{[\varphi(z)-h(z)][\varphi(z)-h(z)]^{\top}\},
\end{eqnarray}
where the infimum is in the sense that
\begin{eqnarray}\label{eq18}
\nonumber&&\E\{[\varphi(z)-{\Pi}_{\mathcal{G}}(\varphi(z))][\varphi(z)-{\Pi}_{\mathcal{G}}
(\varphi(z))]^{\top}\}\\
\leq && \E\{[\varphi(z)-h(z)][\varphi(z)-h(z)]^{\top}\},
\end{eqnarray}
for all $h\in\mathcal{G}$, where for square matrices $\mathbf A$ and $\mathbf  B$, $\mathbf A\leq \mathbf B$ means that $\mathbf A-\mathbf B$ is negative semidefinite.

Denote $\theta(z)=\E[\mathbf{X}_t|\mathbf{Z}=z]$ and $m(z)$ is the projection of $\theta(z)$ onto $\mathcal{G}$, i.e., $m(z)=\E_{\mathcal{G}}[\theta(z)]$. For $t=1,\ldots,T$ we define $\boldsymbol{\xi}_t = \mathbf{X}_t -\boldsymbol{m}(\mathbf{Z})$, $\boldsymbol{\eta}(\mathbf{Z}) = \theta(\mathbf{Z})-\boldsymbol{m}(\mathbf{Z})$, and
$\boldsymbol{\varepsilon}_t= \mathbf{X}_t -\theta(\mathbf{Z})$, where $\boldsymbol{\xi}_t$, $\boldsymbol{\eta}(\mathbf{Z})$, and $\boldsymbol{\varepsilon}_t$ are $N\times Q$ matrices. Also, the following conditions about the data generating process, basis functions, factor loadings, and sieve approximation are required to obtain the large sample properties of the proposed estimator, $\boldsymbol{\boldsymbol{\widehat{\boldsymbol{\beta}}}}$.

\begin{assumption}[Data generating process]\label{asum0}\hfill
\begin{description}
\item[(i)] $\theta(z)$, $m(z)$, and $\eta(z)$ are bounded functions in $\mathcal{Z}$. 
\item[(ii)] $\sup_{z\in \mathbf{\mathcal{Z}}} \E\left(\left. \boldsymbol\varepsilon_t\boldsymbol\varepsilon^{\top}_t\right| \mathbf{Z}=z \right) < C$, for some $C > 0$, $t=1,\ldots,T$.
\item[(iii)] Define
$\widetilde{V}_{\xi} = \operatorname{plim}_{N,T \rightarrow \infty}\frac{1}{NT}\sum_t \boldsymbol\xi^{\top}_t\boldsymbol\xi_t.$
$\widetilde{V}_{\xi}$ is finite and positive definite.
\end{description}
\end{assumption}

\begin{assumption}[Identification]\label{asum2}
Almost surely, $T^{-1}\mathbf F^{\top}\mathbf F=\mathbf I_K$.
\end{assumption}

Assumption \ref{asum0} allows for correlation between $\mathbf{X}_{it}$ and $\mathbf{Z}_{i}$ through $\theta(\mathbf{Z})$, $m(\mathbf{Z})$ and $\eta(\mathbf{Z})$. This assumption is standard in semiparametric estimation techniques (see for example Assumption 2.1(ii) in \citet{AHMADLEELAHANONLI:2005}). Also, Assumption \ref{asum2} is commonly used in the estimation of factor models and enables to identify separately the factors $\mathbf{F}$ (see condition PC1 in \citet{Bai-Ng2013}). Assumption \ref{asum0} (iii) guarantees that  $(NT)^{-1}\sum_{t=1}^T \mathbf{X}_t^{\top} \left[\mathbf{I}_N- \mathbf P_{\Phi}(\mathbf{Z})\right]\mathbf{X}_t$ is asymptotically nonsingular.

\begin{assumption}[Sieve basis functions]\label{asum3}\hfill
\begin{description}
\item[(i)] There are two positive constants, $c_{\min}'$ and $c_{\max}'$ such that, with probability approaching one (as $N\rightarrow\infty$),
    \begin{eqnarray*}
c_{\min}' < \lambda_{min}\left(N^{-1}\Phi(\mathbf{Z})^{\top}\Phi(\mathbf{Z})\right)<
\lambda_{max}\left(N^{-1}\Phi(\mathbf{Z})^{\top}\Phi(\mathbf{Z})\right) <c_{\max}'.
\end{eqnarray*}
\item[(ii)] $\max_{j \le J_N ,i\leq N, d\leq D}\E[\phi_{j}(Z_{id})^2]<\infty$.
\end{description}
\end{assumption}

As already pointed out in \citet{Fan-Liao-Wang2016}, $N^{-1}\Phi(\mathbf{Z})^{\top}\Phi(\mathbf{Z})=N^{-1}\sum^N_{i=1}\phi\left(\mathbf{Z}_i\right)^{\top}\phi\left(\mathbf{Z}_i\right)$ and $\phi\left(\mathbf{Z}_i\right)$ is of order $J_ND$ much smaller than $N$. Thus, condition (i) can follow from a strong law of large numbers. This condition can be satisfied through proper normalizations of commonly used basis functions.

The following set of conditions is concerned with the accuracy of the sieve approximation.

\begin{assumption}[Accuracy of sieve approximation]\label{asum4}\hfill
\begin{description}
\item[(i)] For $k=1,\ldots,K$, $g_k(\cdot) \in \mathcal{G}$ and for $q=1,\ldots,Q$, $m_q(\cdot) \in \mathcal{G}$, where $m_q(\cdot)$ is the $q$th column of $m(\cdot)$.
\item[(ii)] For $k=1,\ldots,K$, $d=1,\ldots,D$, $q=1,\ldots,Q$, and $i=1,\ldots,N$, and let $r$ and $\zeta$ be elements already stated in Definition \ref{definition}. The sieve coefficients $\{b_{j,kd}\}^{J_N}_{j=1}$ and $\{c_{j,qd}\}^{J_N}_{j=1}$ satisfy, for $\kappa=2(r+\zeta)\geq4$ as $J_N\rightarrow\infty$, 
\begin{align*}
    \sup_{z \in \mathcal{Z}_{d}} \left|g_{kd}(z)-\sum_{j=1}^{J_N}b_{j,kd}\phi_{j}(z)\right|^2=\mathcal{O}\left(J^{-\kappa}_N\right),\\
    \sup_{z \in \mathcal{Z}_{d}} \left|m_{qd}(z)-\sum_{j=1}^{J_N}c_{j,qd}\phi_{j}(z)\right|^2=\mathcal{O}\left(J^{-\kappa}_N\right),
\end{align*}
where $m_{qd}(z)$ is the $d$-th additive element of $m_q(z)$, $\mathcal{Z}_{d}$ is the support of the $d$-th element of $\mathcal{Z}$, and $J_N$ is the sieve dimension.
\item[(iii)] $\max_{j,k,d}b_{j,kd}^2<\infty$, $\max_{j,q,d}c_{j,qd}^2<\infty$.
\end{description}
\end{assumption}

As it is remarked in \citet{Fan-Liao-Wang2016}, Assumption \ref{asum4} (ii) is satisfied by the use of common basis functions such as polynomial basis or B-splines. In particular, \citet{Lorentz1986} and \citet{Chen2007} show that (i) implies (ii) in this particular case. 

The next assumption refers to the error factor loadings $\mathbf{\gamma}_i$, for $i=1,\ldots,N$.

\begin{assumption}[Error factor loadings]\label{asum5}\hfill
\begin{description}
\item[(i)] $\{\boldsymbol{\boldsymbol \gamma}_i\}_{i\leq N}$ is independent of $\left\{\mathbf{Z}_{i}\right\}_{i\leq N}$. Furthermore, conditionally on $\mathbf{f}_1,\ldots,\mathbf{f}_T$, $\{\boldsymbol{\boldsymbol \gamma}_i\}_{i\leq N}$ is independent of $\left\{\boldsymbol{\xi}_t\right\}_{t\le T}$ and $\E(\gamma_{ik})=0$ for $k=1,\ldots,K$.
\item[(ii)] $\max_{k\leq K, i\le N} \E\left[g_k(\mathbf{Z}_i)^2\right] <\infty$. Also, $\nu_N<\infty$ and
\[
\max_{k\leq K,j\leq N}\sum_{i\leq N}|\E(\gamma_{ik}\gamma_{jk})|=\mathcal{O}(\nu_N),
\]
where
\[
\nu_N=\max_{k\leq K}N^{-1}\sum_{i\leq N}{\sf Var}(\gamma_{ik}),  
\]
\item[(iii)] For some $\delta >2$,
\begin{equation}
            \max_{i\le N; k\le K} \E\left|\gamma_{ik}\frac{1}{T}\sum^T_{t=1}\E\left(\left.\xi_{itq}f_{kt}\right| \boldsymbol\Gamma\right)\right|^{\delta} < \infty, \quad q=1,\ldots,Q.
        \end{equation}
\end{description}
\end{assumption}

Note that in Assumption 3.5 (ii) we assume cross-sectional dependence of the error factor loadings. To show the consistency of the proposed estimator for simplicity we can assume the independence of the factor loadings $\gamma_{ik}$ from the random part of the covariates, $\textbf{Z}_i$, but we do not need to impose a restrictive i.i.d. assumption.

Through the paper, some regularity conditions about weak dependence and stationarity are assumed on the factors and the idiosyncratic terms. In particular, we impose strong mixing conditions. Let $\mathcal{F}_{-\infty}^0$ and $\mathcal{F}_{T}^{\infty}$ denote the $\sigma$-algebras generated by $\{(\boldsymbol\xi_t,\mathbf{f}_t,\mathbf u_t):t\leq 0\}$ and $\{(\boldsymbol\xi_t,\mathbf{f}_t,\mathbf u_t):t\geq T\}$, respectively. Define the mixing coefficient
\[
\alpha(T)=\sup_{A\in\mathcal{F}_{-\infty}^0,B\in\mathcal{F}_T^{\infty}}|\P(A)\P(B)-\P(AB)|.
\]

\begin{assumption}[Data generating process]\label{asum6}\hfill
\begin{description}
\item[(i)] $\{\boldsymbol\xi_t,\mathbf u_t,\mathbf{f}_t\}_{t\leq T}$ is strictly stationary, $\{\mathbf u_t\}_{t\le T}$ is independent of $\{\mathbf{Z}_i,\boldsymbol{\gamma}_i,\boldsymbol\xi_t,\mathbf{f}_t\}_{i\le N; t\le T}$ and $\E(u_{it})=0$ for all $i\leq N$, $t\leq T$.
        \item[(ii)] For some $\delta >2$,
        \begin{align}
 \max_{t\le T} \E\left|\xi_{itq}u_{it}\right|^{\delta} < \infty, \quad i=1,\ldots,N; \quad q = 1,\ldots,Q,\\
 \max_{t\le T}\max_{k\leq K}\E|\xi_{itq}f_{tk}|^{\delta}=M_{\delta}<\infty,\quad i=1,\ldots,N;\quad q=1,\ldots,Q.
        \end{align}
\item[(iii)] Strong mixing: $\alpha(k) \leq a k^{-\tau}$, where $a$ is a positive constant and $\tau > \frac{\delta}{\delta-2}$.\\

\item[(iv)] Weak dependence: there is $C>0$ so that
\begin{align*}
    \max_{j\leq N}\sum_{i=1}^N|\E(u_{it}u_{jt})|&<C,
    \\
    (NT)^{-1}\sum_{i=1}^N\sum_{j=1}^N\sum_{t=1}^T\sum_{s=1}^T
    |\E(u_{it}u_{js})|&<C, 
    \\
    \max_{i\le N}(NT)^{-1}\sum_{l=1}^N\sum_{l'=1}^N\sum_{t=1}^T\sum_{s=1}^T
    \left|{\sf Cov}\left(u_{it}u_{lt},u_{is}u_{l's}\right)\right| & < C.
\end{align*}
\end{description}
\end{assumption}

Assumption \ref{asum6} is standard in factor analysis \citep{Bai2003,Stock-Watson2002,Fan-Liao-Wang2016}. Part (i) is standard in partially linear models \citep{AHMADLEELAHANONLI:2005, Hardle2000}. The independence assumption between $\mathbf u_t$ and $\{\mathbf{Z}_i, \boldsymbol\xi_t\}$ can be relaxed by allowing for conditional independence. Part (iii) is a strong mixing condition for the weak temporal dependence of $\{\boldsymbol\xi_t,\mathbf u_t,\mathbf{f}_t\}$, whereas (iv) imposes weak cross-sectional dependence in $\{u_{it}\}_{i\leq N, t\leq T}$. This condition is usually satisfied when the covariance matrix of the error term $u_{it}$ is sufficiently sparse under the strong mixing condition and it is commonly imposed for high-dimensional factor analysis. 

\subsection{Limiting Theory}
A very intuitive idea of the asymptotic behavior of our estimator can be obtained by plugging (\ref{eq4}) in (\ref{eq12}) that yields 
\begin{align*}
\boldsymbol{\boldsymbol{\widehat{\boldsymbol{\beta}}}}-\boldsymbol{\beta}&=\left[\sum_t\mathbf{X}_t^{\top}\mathbf M_{\Phi}(\mathbf{Z})
\mathbf{X}_t\right]^{-1}\sum_t\mathbf{X}_t^{\top}\mathbf M_{\Phi}(\mathbf{Z})(\boldsymbol \Lambda \mathbf{f}_t+\mathbf u_t),
\end{align*}
where $\mathbf M_{\Phi}(\mathbf{Z})=\mathbf{I}_N-\mathbf{P}_{\Phi}(\mathbf{Z})$ and $\mathbf{\Lambda} = \mathbf{G(Z)}+\mathbf{\Gamma}$. As the reader can see in the above expression, there is a direct dependence of $\boldsymbol{\widehat{\boldsymbol{\beta}}}$ on the unobserved factor loadings through $(NT)^{-1}\sum_t\mathbf{X}_t^{\top}\mathbf M_{\Phi}(\mathbf{Z})\boldsymbol \Lambda \mathbf{f}_t$. Nevertheless, using (\ref{eq3}) and given that it can be proved that $(NT)^{-1}\sum_t\mathbf{X}_t^{\top}\mathbf M_{\Phi}(\mathbf{Z})\mathbf{G}(\mathbf{Z})\mathbf{f}_t={\scriptstyle\mathcal{O}}_p(1/\sqrt{NT})$ (see the proof of Theorem 3.1 in the Supplementary Material document), we have that
\begin{align}\label{aaa1}
\boldsymbol{\boldsymbol{\widehat{\boldsymbol{\beta}}}}-\boldsymbol{\boldsymbol{\beta}} =\left[\sum_t\mathbf{X}_t^{\top}\mathbf M_{\Phi}(\mathbf{Z})
\mathbf{X}_t\right]^{-1}\sum_t\mathbf{X}_t^{\top}\mathbf M_{\Phi}(\mathbf{Z})(\boldsymbol \Gamma \mathbf{f}_t+\mathbf u_t)+{\scriptstyle\mathcal{O}}_p\left(\frac{1}{\sqrt{NT}}\right).
\end{align}
In this situation, we can conclude that the limiting distribution of $\boldsymbol{\boldsymbol{\widehat{\boldsymbol{\beta}}}}-\boldsymbol{\boldsymbol{\beta}}$ only depends on idiosyncratic terms (related to both the error term and the approximation error of the basis functions to the factor loadings). Under Assumptions \ref{asum0}--\ref{asum6} it is also possible to show (see the Appendix A of the Supplementary Material document for a related proof) that, as both $N$ and $T$ tend to infinity,
\begin{equation*}\label{aa1}
    \boldsymbol{\boldsymbol{\widehat{\boldsymbol{\beta}}}} -\boldsymbol{\beta} = \mathcal{O}_p\left(1/\sqrt{NT}\right) + \mathcal{O}_p\left(\sqrt{\nu_N/N}\right). 
\end{equation*}
The interesting feature of this asymptotic bound is that the rate of convergence of $\widehat{\boldsymbol\beta}$ can be slower than $\sqrt{NT}$ depending on the behavior of $\nu_N$ (i.e. the variance of the error factor loadings).
In order to clarify this, we further take a look at the asymptotic distribution of our estimator. The previous result on the consistency and the convergence rate is based on weak dependence in the error term and idiosyncratic factor loadings. To show asymptotic normality, we have to impose the stronger condition of cross-sectional independence while still allowing for weak dependence in the time dimension.

\begin{assumption}\label{asum66}
$\left\{\boldsymbol{u}_i,\boldsymbol{\gamma}_{i}\right\}_{i\le N}$ are independent and non-identically distributed random variables across $i$.
\end{assumption}

Then, the asymptotic distribution of the projection-based interactive fixed effects estimator $\boldsymbol{\boldsymbol{\widehat{\boldsymbol{\beta}}}}$ is provided in the following theorem.

\begin{theorem}[Limiting distribution]\label{theo1}
Under assumptions \ref{asum0}--\ref{asum66} and if it is further assumed that, for $\kappa\geq4$ and $\varrho \in (\frac{1}{\kappa},\frac{1}{2})$, 
$J_N \sim N^{\varrho}$
and $T/N^{\kappa\varrho-1}\rightarrow 0$, as both $N$ and $T$ tend to infinity, then
for $\vartheta \in [0,1)$,
\begin{equation}
\sqrt{NT^{\vartheta}}(\boldsymbol{\widehat{\beta}}-\boldsymbol{\beta})\stackrel{\mathcal{L}}{\rightarrow} N\left(0,\widetilde{V}^{-1}_{\xi}
\widetilde{V}_{\Gamma}\widetilde{V}^{-1}_{\xi}\right).
\end{equation}
Under the same set of assumptions, if $\vartheta = 1$
\begin{equation}
\sqrt{NT}(\boldsymbol{\boldsymbol{\widehat{\boldsymbol{\beta}}}}-\boldsymbol{\beta})\stackrel{\mathcal{L}}{\rightarrow} N\left(0,\widetilde{V}^{-1}_{\xi}
\left(\widetilde{V}_{\Gamma}+ \widetilde{V}_{u}\right)\widetilde{V}^{-1}_{\xi}\right),
\end{equation}
and finally, if $\vartheta > 1$
\begin{equation}
\sqrt{NT}(\widehat{\boldsymbol{\beta}}-\boldsymbol{\beta})\stackrel{\mathcal{L}}{\rightarrow} N\left(0,\widetilde{V}^{-1}_{\xi}
\widetilde{V}_{u}\widetilde{V}^{-1}_{\xi}\right),
\end{equation}
where
\begin{eqnarray}
\widetilde{V}_{\xi} &  \stackrel{\operatorname{def}}{=} & \operatorname{plim}_{N,T \rightarrow \infty} \frac{1}{NT}\sum^T_{t=1}{\boldsymbol\xi^{\top}_t\boldsymbol\xi_t}, \\
 \widetilde{V}_{\Gamma} & \stackrel{\operatorname{def}}{=} & \operatorname{lim}_{N,T\to\infty}\frac{1}{N}\sum^N_{i=1}\sum_{k\le K; k^{\prime} \le K} \E\left[\gamma_{ik}\gamma_{ik^{\prime}}\frac{1}{T^2}\sum_{t,s}\E\left(\left.\boldsymbol{\xi}_{it}f_{tk}\right| \boldsymbol \Gamma\right)\E\left(\left.\boldsymbol{\xi}_{is}f_{sk^{\prime}}\right| \boldsymbol \Gamma\right)^{\top}\right], \\
 \widetilde{V}_{u} & \stackrel{\operatorname{def}}{=} & \operatorname{lim}_{N,T\to\infty}\frac{1}{NT}\sum_{t=1}^{T}\sum^{T}_{t'=1 }\E(\boldsymbol\xi_t^{\top}
\mathbf u_t\mathbf u^{\top}_{t'}\boldsymbol \xi_{t'}).
\end{eqnarray}
\end{theorem}

The proof of Theorem \ref{theo1} is provided in Appendix B.1 in the Supplementary Material document. The key component of the proof is following the Frisch-Waugh Theorem to partial-out the effect of the latent factors and corresponding loadings. We want to highlight that the relative rate requirements of $N$ and $T$ crucially depend on the smoothness parameter, $\kappa$. In particular, if $\kappa=4$ we have the requirement that $T/N$ tends to zero regardless of the choice for the sieve dimension $\varrho$. The constraints in the rates of growth imposed on $N$ and $T$ are similar to other assumptions used in similar literature such as in \citet{AHMADLEELAHANONLI:2005}.
Note that Assumption \ref{asum66} is introduced for the sake of simplicity. It is indeed used for the application of the corresponding central limit theorems (CLT) in the proof but it could be relaxed at the cost of a much cumbersome proof.  

\begin{remark}
As we can observe from (\ref{aaa1}) and
Theorem \ref{theo1} the asymptotic distribution of $\boldsymbol{\boldsymbol{\widehat{\boldsymbol{\beta}}}}$ depends on interplay of two leading terms,
\[
\sum_t\mathbf{X}_t^{\top}\mathbf M_{\Phi}(\mathbf{Z})\boldsymbol \Gamma \mathbf{f}_t+\sum_t\mathbf{X}_t^{\top}\mathbf M_{\Phi}(\mathbf{Z})\mathbf u_t.
\]

The term $(NT)^{-1}\sum_t\mathbf X^{\top}_t\mathbf M_{\Phi}(\mathbf{Z})\boldsymbol \Gamma \mathbf{f}_t$ arises from the cross-sectional estimation, and it shows a rate of order $\mathcal{O}_p\left(N^{-1/2}\sqrt{\nu_N}\right)$ and the term $(NT)^{-1}\sum_t\mathbf X^{\top}_t\mathbf M_{\Phi}(\mathbf{Z})\mathbf u_t$ which has a leading term of order $\mathcal{O}_p\left((NT)^{-1/2}\right)$ (see \textbf{Proof of Theorem 3.1 (iii)} of the Supplementary Material). Indeed, the interaction between these two leading terms affects crucially the resulting rate of convergence of the limiting distribution. As it can be observed in Theorem \ref{theo1}, this rate is affected by the behavior of $\nu_N$. This term reflects the strength of the relationship between the $\boldsymbol{\boldsymbol \lambda}_i$'s and the $\mathbf{Z}_{i}$'s. When a relevant part of the variation of the loading coefficients $\boldsymbol{\boldsymbol \lambda}_i$ is explained by $\mathbf{Z}_i$ (that is $\nu_N$ is close to zero) the observed characteristics capture almost all fluctuations of $\widehat{\boldsymbol{\beta}}-\boldsymbol{\beta}$, leading to a faster rate of convergence $\sqrt{NT}$. On the other hand, if $\nu_N$ is far from zero, then the fluctuations of $\widehat{\boldsymbol{\beta}}-\boldsymbol{\beta}$ can be explained mostly by cross-sectional variation, and therefore time series regression is not relevant to help to remove the correlation between loading coefficients and covariates when estimating the $\boldsymbol{\beta}$'s. In this case, the limiting distribution is determined by a cross-sectional CLT and hence the rate of convergence is slower (note that for $\nu_N = \mathcal{O}(1)$ the rate is $\sqrt{N}$).
\end{remark}

\begin{remark}From Theorem \ref{theo1} it is also possible to identify specifications under which our estimator might outperform the PCA or the CCE estimators. If $\boldsymbol{\lambda}_i = \mathbf{g(Z_i)} + \boldsymbol\gamma_i$ and $\nu_N \approx 0$ our estimator appears as more efficient as the others. If we further assume that the latent factor loadings can be completely explained by the nonparametric functions, i.e., $\boldsymbol \Gamma=0$ as in \citet{Zhang_Zhou_Wang2021}, and given the idiosyncratic error terms are i.i.d. with ${\sf Var}(u_{it})=\sigma^2$, our estimator is semiparametrically efficient in the sense that the inverse of the asymptotic variance of
$\sqrt{NT}(\boldsymbol{\boldsymbol{\widehat{\boldsymbol{\beta}}}}-\boldsymbol{\beta})$ equals the semiparametric efficiency bound. From the result of \citet{chamberlain1992efficiency} the semiparametric efficiency bound for the inverse of the asymptotic variance of an estimator of $\boldsymbol{\beta}$ is
\begin{align}\label{eeq1}
    \mathcal{J}_{0}= \operatorname{inf}_{g\in \mathcal{G}}\E\left\{\left[\mathbf{X}_{it}-\mathbf g(\mathbf{Z}_i)\right]
    {\sf Var}\left(u_{it}\right)^{-1}\left[\mathbf{X}_{it}-\mathbf g(\mathbf{Z}_i)\right]^{\top}\right\}.
\end{align}

Under the i.i.d. Assumption, (\ref{eeq1}) can be rewritten as
\begin{eqnarray*}
    \mathcal{J}_{0} & = & \frac{1}{\sigma^2}\operatorname{inf}_{g\in \mathcal{G}}\E\left\{\left[\mathbf{X}_{it}-\mathbf g(\mathbf{Z}_i)\right] \left[\mathbf{X}_{it}-\mathbf g(\mathbf{Z}_i)\right]^{\top}\right\} \\
    & = & \frac{1}{\sigma^2}\E\left\{\left[\mathbf{X}_{it}-\mathbf m(\mathbf{Z}_i)\right] \left[\mathbf{X}_{it}- \mathbf m(\mathbf{Z}_i)\right]^{\top}\right\} \\
    & = & \frac{1}{\sigma^2}\E\left\{\boldsymbol{\xi}_{it}\boldsymbol{\xi}^{\top}_{it}\right\}.
\end{eqnarray*}

Note that the inverse of the last expression coincides with the asymptotic variance of $\sqrt{NT}(\boldsymbol{\widehat{\boldsymbol{\beta}}}-\boldsymbol{\beta})$ when the error terms are uncorrelated and homoskedastic. Then, $\boldsymbol{\widehat{\boldsymbol{\beta}}}$ is a semiparametrically efficient estimator under these assumptions.
\end{remark}

\begin{remark}
It is possible to estimate the latent factors and loading coefficients from the regression residuals using the Projected-PCA method of \citet{Fan-Liao-Wang2016}. Let $\mathbf{\widetilde{y}}_{t}=\mathbf y_{t}-\mathbf X_{t}\boldsymbol{\boldsymbol{\widehat{\boldsymbol{\beta}}}}$,
and let $\mathbf{\widetilde{Y}}=(\mathbf{\widetilde{y}}_{1},\ldots,\mathbf{\widetilde{y}}_{T})$. Now the matrix of factors $\mathbf F$ and $\mathbf{G}(Z)$ can be recovered from the projected matrix of residuals $\mathbf{P}_{\Phi}(\mathbf{Z})\mathbf{\widetilde{Y}}$. The asymptotic properties and the resulting convergence rates remain unaffected by the need to estimate the regression coefficients in a first step. We provide details on the estimation of the latent factors and loadings in Section C of the Supplementary Material document.
\end{remark}

\subsection{Uniformly Valid Inference via the Cross-Sectional Bootstrap}
The results of Theorem \ref{theo1} have important implications for conducting inference on the estimated regression parameters. In particular, since the variance of the idiosyncratic part of the factor loadings decides which of the two terms will be the leading one, it will ultimately determine the convergence rate of our estimator. As a consequence, the asymptotic distribution of $\boldsymbol{\widehat{\boldsymbol{\beta}}}-\boldsymbol{\beta}$ has a discontinuity when the variance of the factor loadings is close to the boundary. In financial econometrics studies, this issue is often circumvented by assuming weak heterogeneity as a default setting \citep{Connor-Linton2007,connor2012efficient}. However, recent studies such as \citet{Fan-Liao-Wang2016} find empirical evidence for the case of strong heterogeneity. Further, usual plug-in approaches based on estimated asymptotic covariance matrices will lead to misleading conclusions if $\nu_N=\mathcal{O}(T^{-1})$, as they provide confidence intervals that are too wide because the asymptotic covariance matrix is over-estimated, and the coverage probabilities will be too conservative \citep{liao2018uniform,FernndezVal2022DynamicHD}. Similarly, simply ignoring the cross-sectional term will lead to under-coverage in the strong heterogeneity case. 

Uniformly valid inference for panel data models is an important topic beyond the specifics of our model setup. For instance, \citet{lu2023uniform} observe a similar issue in a panel model with two-dimensional heterogeneity in the regression parameters. In their case, the issue is caused by the level of temporal and cross-sectional heterogeneity in the slope coefficients.

Fortunately, the uniformity issue can be solved by using the cross-sectional bootstrap proposed by \citet{kapetanios2008bootstrap}. Besides achieving uniformly valid inference, the approach is both intuitive and easy to implement. The basic idea is to sample with replacement cross-sectional units while keeping the entire time series of the sampled individual units unchanged. By doing this, the resampling scheme directly mimics the cross-sectional variations in $\boldsymbol \Gamma$, regardless of the underlying level of heterogeneity. The consequence is a uniformly valid inference. This is in direct contrast to \citet{andrews2000inconsistency}, who found that the usual bootstrap will lead to inconsistency when a parameter is on the boundary of the support. The reason why this problem does not occur in our case is that we do not explicitly model the variance of the idiosyncratic factor loadings as a parameter, i.e., it does not appear in the loss function of our least squares problem. 

A crucial assumption for the bootstrap validity is that the data is cross-sectionally independent. In fact, \citet{menzel2021bootstrap} showed that uniform non-conservative inference is impossible under general dependence in more than one dimension. Recently, \citet{de2024cross} studied the theoretical properties of the cross-sectional bootstrap for the CCE approach of \citet{Pesaran2006} and proposed a bias-correction procedure in the asymptotic regime $N/T\to\rho<\infty$. The uniform validity of the bootstrap procedure in settings similar to ours was recently shown in \citet{liao2018uniform} and \citet{FernndezVal2022DynamicHD}. The specific aspect of our procedure is that we only resample cross-sectional units after projecting the data, i.e., removing the effect of $\boldsymbol{Z}_i$ on the factor loadings.

As a positive side effect, the cross-sectional bootstrap is able to keep the dependence in the time dimension. Therefore, the inference is also robust towards serial dependence in the idiosyncratic error term and in the latent factors. In the following, we summarize the steps of the cross-sectional bootstrap procedure.
\begin{enumerate}[Step 1:]
    \item Choose a confidence level $\alpha$, and the number of bootstrap samples, $B$.
    \item Regress $y_{it}$ and $X_{itq}$ on $\Phi(\mathbf{Z})$, and obtain residuals, $\mathbf{\dot{y}}_t\stackrel{\text{def}}{=}[\mathbf{I}_N-\mathbf P_{\Phi}(\mathbf{Z})]\mathbf y_t$ and $\mathbf{\dot{X}}_t\stackrel{\text{def}}{=}[\mathbf{I}_N-\mathbf P_{\Phi}(\mathbf{Z})]\mathbf{X}_t$, $t=1,\ldots,T$.
    \item Calculate $\boldsymbol{\widehat{\boldsymbol{\beta}}}=(\sum_{t=1}^T\mathbf{\dot{X}}_t^\top\mathbf{\dot{X}}_t)^{-1}\sum_{t=1}^T\mathbf{\dot{X}}_t^\top \mathbf{\dot{y}}_t$.
    \item For $b=1,\ldots,B$, draw a sample of $N$ cross-sectional units with replacement while keeping the unit's entire time series unchanged. Denote the resulting matrices of regressors and vectors of dependent variables by $\mathbf{\dot{X}}_{b,t}^*$ and $\mathbf{\dot{y}}_{b,t}^*$, respectively.
    \item Obtain the bootstrap estimate $\boldsymbol{\widehat{\boldsymbol{\beta}}}_b^*=(\sum_{t=1}^T\mathbf{\dot{X}}_{b,t}^{*\top}\mathbf{\dot{X}}^*_{b,t})^{-1}\sum_{t=1}^T\mathbf{\dot{X}}_{b,t}^{*\top} \mathbf{\dot{y}}_{b,t}^*$.
    \item Calculate the ($1-\alpha$)-confidence interval for the $j$-th component of $\boldsymbol{\beta}_{j}$,
    \begin{align*}
        CI_{\alpha}(\boldsymbol{\beta}_j)=\boldsymbol{\widehat{\boldsymbol{\beta}}}_j\pm q_{\alpha,j},
    \end{align*}
    where $q_{\alpha,j}$ is the $(1-\alpha)$-quantile of the bootstrap distribution of $|\boldsymbol{\widehat{\boldsymbol{\beta}}}_{b,j}^*-\boldsymbol{\widehat{\boldsymbol{\beta}}}|$. Or, more generally, for $v\in\mathbb{R}^Q$,
    \begin{align*}
        CI_\alpha(v^\top\boldsymbol{\beta})=v^{\top}\boldsymbol{\widehat{\boldsymbol{\beta}}}\pm q_{\alpha,v},
    \end{align*}
    where $q_{\alpha,v}$ is the $(1-\alpha)$-quantile of the bootstrap distribution of $|v^{\top}(\boldsymbol{\widehat{\boldsymbol{\beta}}}^*_b-\boldsymbol{\widehat{\boldsymbol{\beta}}})|$.
\end{enumerate}

For the bootstrap validity, we need to assume the existence of a consistent estimator of the variance of $v^{\top}\widehat{\boldsymbol{\beta}}$.
\begin{assumption}\label{assum:covariance}
    Denote $V_{\boldsymbol{\beta},v}=\lim_{N,T\to\infty}{\sf Var}[\sqrt{NT^{\vartheta}}v^{\top}(\widehat{\boldsymbol{\beta}}-\boldsymbol{\beta})]$, if $\vartheta \in [0,1)$ and $V_{\boldsymbol{\beta},v}=\lim_{N,T\to\infty}{\sf Var}[\sqrt{NT}v^{\top}(\widehat{\boldsymbol{\beta}}-\boldsymbol{\beta})]$, if $\vartheta \ge 1$. There exists a consistent estimator $V_{\boldsymbol{\beta},v,n}$, satisfying $V_{\boldsymbol{\beta},v}^{-1/2}-V_{\boldsymbol{\beta},v,n}^{-1/2}=\Co_p(1)$.
\end{assumption}

The following theorem provides the bootstrap validity, uniformly over settings with varying degrees of variability in the idiosyncratic factor loadings.

\begin{theorem}[Bootstrap Validity]\label{theo_bootstrap}
Let $\{\P_T:T\geq1\}\subset \mathcal{P}$ be sequences of probability laws. Let the conditions of our Theorem \ref{theo1} hold uniformly over these sequences. Further assume that $u_{it}$ and $\boldsymbol{\boldsymbol \gamma}_i$ are cross-sectionally independent. Then we have, uniformly for all $\{\P_T:T\geq1\}\subset\mathcal{P}$, and for a confidence level $1-\alpha$,
\begin{align*}
    \P_T\left(v^\top\boldsymbol{\beta}\in CI_{\alpha}(v^\top\beta)\right)\to1-\alpha.
\end{align*}
\end{theorem}

The proof of Theorem \ref{theo_bootstrap} can be found in Appendix B.2 in the Supplementary Material document. An essential part of the proof is to show that the asymptotic expansion of the bootstrap version of the estimator is identical to that of the original estimator.

\section{Numerical Studies}\label{sec4}
In this section, we evaluate the finite-sample performance of our estimator in a simulation study. We are interested both in the estimation accuracy of the parameter vector, $\boldsymbol{\beta}$, and the empirical coverage probabilities of the cross-sectional bootstrap procedure. Throughout the study, we fix the number of factors, $K=3$, the dimension of the time-invariant variable is set to $D=2$, and the dimension of covariates is set to $Q=2$. The true regression coefficients are $\boldsymbol{\beta}=(2,-1)^{\top}$. The time-invariant variables are generated by i.i.d. $Z_{id}\sim U[-1,1]$. The covariates are generated by setting $X_{itq}=\mathbf a_{iq}^\top \mathbf{f}_t+2(\sqrt{ g_1(\mathbf{Z}_{i})},\ldots,\sqrt{g_K(\mathbf{Z}_{i})})^{\top}\boldsymbol{b}_q+\pi_{itq}$, where $\pi_{itq}\sim N(0,1)$ i.i.d., $a_{iqk}\sim U[-0.5,0.5]$ and $b_{qk}\sim U[-1,1]$. We generate the latent factors, $(f_{k1},\ldots,f_{kT})$, as MA$(\infty)$ processes with algebraic decay and under independence across factors for all $k$. The factor loadings are set to $\lambda_{ik}=g_{k}(\mathbf{Z}_{i})+\gamma_{ik}$, where
$g_1(z)=\sin(2z_1)^3+\cos(z_2^2)$, $g_2(z)=-\tan(z_1^2)+2\cos(z_2+1)$ and $g_3(z)=z_2^3-\sin(3z_1)$.

Finally, for the idiosyncratic error term, we consider the case of i.i.d. standard normal $u_{it}$ as well as the case of weak temporal dependence, in which $(u_{i1},\ldots,u_{iT})$ are generated from a MA$(\infty)$ process with algebraic decay parameter $5$. For the idiosyncratic part of the factor loadings we consider three settings. First, in the strong heterogeneity case $\boldsymbol\gamma_{i}\sim N(0,0.5)$ (i.e., $\nu_N=\mathcal{O}(1)$). Second, we consider the special case $\nu_N=0$. Third, we consider the weak factor case in which $\boldsymbol\lambda_i^{\top}\mathbf{f}_t=\mathcal{O}(T^{-1/2})$ (i.e., $\nu_N=\mathcal{O}(T^{-1})$). In this numerical study, we rely on B-spline basis functions and we select $J_N=\lceil N^{1/3}1.5\rceil$. For each setting $500$ Monte Carlo runs are conducted.

We compare the performance of our projection-based interactive fixed effects (P-IFE) estimator for $\boldsymbol{\beta}$ with the principal component-based interactive fixed effects (PC-IFE) estimator of \citet{Bai2009} and a bias-corrected version of the same estimator (bc-PC-IFE). For these comparisons, we rely on the R package \texttt{phtt} \citep{bada2014phtt}. The number of factors is selected according to the PC1 criterion in \citet{Bai-Ng2002}. As performance measures, we consider the root mean square error ($RMSE$).
The simulation results under Gaussian disturbances for different values of $\nu_N$, $N$, and $T$ are reported in Table \ref{table:rmse_normal}. The $RMSE$ of our P-IFE can be effectively reduced with increasing sample size. For the strong heterogeneity case, we observe an advantage of the PC-IFE for small and medium samples. For $N=500$ this advantage is reversed and the P-IFE has a higher accuracy. In the other two settings for $\nu_N$, we can see that the P-IFE outperforms its competitors in almost all cases. The outperformance is best visible for settings with large sample sizes in the case of $\nu_N=0$. In particular, for $N=500$ the $RMSE$ of our P-IFE is less than a third of that of the PC-IFE.

The results for serially dependent error terms are displayed in Table \ref{table:rmse_ar}. Again, we can observe that the PC-IFE outperforms the P-IFE in $\nu_N=\mathcal{O}(1)$ case for small and medium sample sizes. Also similar to the i.i.d. case, the P-IFE has the lowest $RMSE$ in all settings for $\nu_N=0$ and $\nu_N=\mathcal{O}(T^{-1})$. Interestingly, the bias-corrected estimator performs worse than the estimator without bias correction. As a robustness check, we also consider $t$-distributed error terms in the Supplementary Material document. See Table \ref{table:rmse_t} for the results. We also consider a more complicated additional data-generating process with a larger number of factors, $K=10$. The results are displayed in Table \ref{table:rmse_factor} for i.i.d. errors and Table \ref{table:rmse_factor_AR} for serially correlated errors. The most notable difference to the results of the first DGP is that the P-IFE outperforms the PC-IFE even in the strong heterogeneity case for settings with a small time dimension, $T=10$. Again, the P-IFE dominates in all settings for $\nu_N=0$ and $\nu_N=\mathcal{O}(T^{-1})$.

\begin{table}[tbp]
\spacingset{1.2} 
\centering
\begin{tabular}{lrr|rrr|rrr}
	\hline \hline
  \multicolumn{2}{r}{} & \multicolumn{3}{c}{$\boldsymbol{\beta}_1$} & \multicolumn{3}{c}{$\boldsymbol{\beta}_2$} \\
  $\nu_N$ & $N$ & $T$ & P-IFE & PC-IFE & bc-PC-IFE & P-IFE & PC-IFE & bc-PC-IFE \\
  \hline
 $\nu_N=\mathcal{O}(1)$ & 50 & 10 & 0.0670 & 0.0582 & 0.0583 & 0.0690 & 0.0605 & 0.0605 \\ 
  & 100 & 10 & 0.0440 & 0.0437 & 0.0439 & 0.0449 & 0.0446 & 0.0448 \\ 
  & 50 & 50 & 0.0385 & 0.0234 & 0.0235 & 0.0399 & 0.0254 & 0.0256 \\ 
  & 100 & 50 & 0.0278 & 0.0195 & 0.0198 & 0.0282 & 0.0204 & 0.0206 \\ 
  & 200 & 100 & 0.0161 & 0.0142 & 0.0146 & 0.0171 & 0.0139 & 0.0139 \\ 
  & 500 & 100 & 0.0100 & 0.0117 & 0.0121 & 0.0100 & 0.0118 & 0.0120 \\ 
  \hline
  $\nu_N=0$ & 50 & 10 & 0.0401 & 0.0569 & 0.0571 & 0.0452 & 0.0581 & 0.0581 \\ 
  & 100 & 10 & 0.0286 & 0.0438 & 0.0440 & 0.0292 & 0.0436 & 0.0437 \\ 
  & 50 & 50 & 0.0205 & 0.0249 & 0.0251 & 0.0196 & 0.0243 & 0.0242 \\ 
  & 100 & 50 & 0.0124 & 0.0208 & 0.0208 & 0.0138 & 0.0210 & 0.0211 \\ 
  & 200 & 100 & 0.0065 & 0.0148 & 0.0150 & 0.0065 & 0.0138 & 0.0138 \\ 
  & 500 & 100 & 0.0038 & 0.0136 & 0.0138 & 0.0039 & 0.0139 & 0.0140 \\ 
  \hline
  $\nu_N=\mathcal{O}(T^{-1})$ & 50 & 10 & 0.0468 & 0.0585 & 0.0585 & 0.0458 & 0.0538 & 0.0539 \\ 
  & 100 & 10 & 0.0319 & 0.0380 & 0.0381 & 0.0308 & 0.0381 & 0.0381 \\ 
  & 50 & 50 & 0.0193 & 0.0197 & 0.0199 & 0.0200 & 0.0196 & 0.0198 \\ 
  & 100 & 50 & 0.0139 & 0.0150 & 0.0154 & 0.0138 & 0.0154 & 0.0154 \\ 
  & 200 & 100 & 0.0065 & 0.0078 & 0.0082 & 0.0066 & 0.0077 & 0.0081 \\ 
  & 500 & 100 & 0.0039 & 0.0058 & 0.0062 & 0.0041 & 0.0057 & 0.0062 \\ 
 \hline \hline
\end{tabular}
\caption{$RMSE$ of the projected IFE estimator, the PC-based IFE, and the bias-corrected PC-based IFE estimator under i.i.d. Gaussian error terms.}
\label{table:rmse_normal}
\end{table}

\begin{table}[tbp]
\spacingset{1.2} 
\centering
\begin{tabular}{lrr|rrr|rrr}
	\hline \hline
  \multicolumn{2}{r}{} & \multicolumn{3}{c}{$\boldsymbol{\beta}_1$} & \multicolumn{3}{c}{$\boldsymbol{\beta}_2$} \\
  $\nu_N$ & $N$ & $T$ & P-IFE & PC-IFE & bc-PC-IFE & P-IFE & PC-IFE & bc-PC-IFE \\
  \hline
 $\nu_N=\mathcal{O}(1)$ & 50 & 10 & 0.0704 & 0.0629 & 0.0629 & 0.0648 & 0.0625 & 0.0626 \\ 
  & 100 & 10 & 0.0438 & 0.0419 & 0.0423 & 0.0463 & 0.0441 & 0.0442 \\ 
  & 50 & 50 & 0.0403 & 0.0249 & 0.0252 & 0.0412 & 0.0252 & 0.0252 \\ 
  & 100 & 50 & 0.0286 & 0.0186 & 0.0189 & 0.0285 & 0.0200 & 0.0201 \\ 
  & 200 & 100 & 0.0160 & 0.0121 & 0.0124 & 0.0165 & 0.0125 & 0.0126 \\ 
  & 500 & 100 & 0.0102 & 0.0108 & 0.0112 & 0.0101 & 0.0112 & 0.0114 \\ 
  \hline
  $\nu_N=0$ & 50 & 10 & 0.0444 & 0.0601 & 0.0604 & 0.0435 & 0.0585 & 0.0586 \\ 
  & 100 & 10 & 0.0297 & 0.0456 & 0.0456 & 0.0288 & 0.0450 & 0.0451 \\ 
  & 50 & 50 & 0.0196 & 0.0258 & 0.0261 & 0.0202 & 0.0259 & 0.0262 \\ 
  & 100 & 50 & 0.0124 & 0.0212 & 0.0211 & 0.0132 & 0.0216 & 0.0218 \\ 
  & 200 & 100 & 0.0064 & 0.0147 & 0.0152 & 0.0063 & 0.0149 & 0.0151 \\ 
  & 500 & 100 & 0.0040 & 0.0129 & 0.0131 & 0.0040 & 0.0133 & 0.0134 \\ 
  \hline
  $\nu_N=\mathcal{O}(T^{-1})$ & 50 & 10 & 0.0463 & 0.0577 & 0.0578 & 0.0455 & 0.0562 & 0.0560 \\ 
  & 100 & 10 & 0.0324 & 0.0375 & 0.0374 & 0.0286 & 0.0364 & 0.0365 \\ 
  & 50 & 50 & 0.0199 & 0.0189 & 0.0191 & 0.0204 & 0.0188 & 0.0190 \\ 
  & 100 & 50 & 0.0138 & 0.0154 & 0.0157 & 0.0136 & 0.0153 & 0.0153 \\ 
  & 200 & 100 & 0.0065 & 0.0077 & 0.0083 & 0.0070 & 0.0084 & 0.0087 \\ 
  & 500 & 100 & 0.0040 & 0.0061 & 0.0066 & 0.0040 & 0.0060 & 0.0063 \\ 
 \hline \hline
\end{tabular}
\caption{$RMSE$ of the projected IFE estimator, the PC-based IFE, and the bias-corrected PC-based IFE estimator under auto-correlated Gaussian error terms (MA($\infty$) with algebraic decay parameter $5$).}
\label{table:rmse_ar}
\end{table}

In the following, we look at the performance of the cross-sectional bootstrap procedure and show its validity in finite samples. As a comparison, we look at the empirical coverage of the PC-IFE estimator. By Corollary 1 in \citet{Bai2009}, under the assumption of i.i.d. error terms, $\sqrt{NT}(\boldsymbol{\widehat{\boldsymbol{\beta}}}_{\text{PC-IFE}}-\boldsymbol{\beta})\stackrel{\mathcal{L}}{\rightarrow}N(0,\sigma^2\mathbf D^{-1})$, where $\mathbf D=\text{plim}(NT)^{-1}\sum_{i=1}^N \mathbf Z_i^\top \mathbf Z_i$, $\mathbf Z_i=\mathbf M_F\mathbf X_i-1/N \sum_{k=1}^N\mathbf M_F\mathbf X_ka_{ik}$, $\mathbf M_F=\mathbf{I}_N-\mathbf F\mathbf F^\top/T$ and $a_{ik}= \boldsymbol\lambda_i^\top(\boldsymbol \Lambda^\top\boldsymbol \Lambda)^{-1} \boldsymbol\lambda_k$. We construct confidence intervals based on the asymptotic distribution with an estimated covariance matrix based on estimated factors and factor loadings.

Table \ref{table:coverage_compare_strong} shows that the empirical coverage of our cross-sectional bootstrap procedure approaches the nominal coverage level as $N$ and $T$ increase. As is often the case, we can observe slight under-coverage in small samples. However, the issue becomes virtually absent in settings with the largest sample size. We want to highlight that these findings hold for all settings for the variance of the idiosyncratic factor loadings, $\nu_N$. We have thus provided evidence for the uniform validity of the bootstrap procedure in finite samples. In the Supplementary Material document, we show that the cross-sectional bootstrap procedure is also robust towards $t$-distributed errors and serially correlated error terms. See Tables \ref{table:coverage_t} and \ref{table:coverage_AR}. Our uniform bootstrap procedure naturally adapts to the data, particularly to potential serial dependence and a varying degree of heterogeneity in the factor loadings.

Looking at the coverage of the asymptotic distribution of the PC-IFE estimator, we can observe under-coverage in all settings. Moreover, the coverage does not improve with increasing sample size. On the contrary, the coverage is worst for the setting with $N=500$ and $T=100$. 

\begin{table}[tbp]
\spacingset{1.2} 
\centering
\begin{tabular}{lrr|rrr|rrr}
	\hline \hline
  \multicolumn{2}{r}{} & \multicolumn{3}{c}{P-IFE} & \multicolumn{3}{c}{PC-IFE} \\
  $\nu_N$ & $N$ & $T$ & $90\%$ & $95\%$ & $99\%$ & $90\%$ & $95\%$ & $99\%$ \\
  \hline
  $\nu_N=\mathcal{O}(1)$ & 50 & 10 & 0.856 & 0.902 & 0.966 & 0.846 & 0.898 & 0.974 \\ 
  & 100 & 10 & 0.856 & 0.932 & 0.982 & 0.836 & 0.902 & 0.970 \\ 
  & 50 & 50 & 0.880 & 0.922 & 0.984 & 0.810 & 0.870 & 0.950 \\ 
  & 100 & 50 & 0.860 & 0.922 & 0.978 & 0.726 & 0.804 & 0.914 \\ 
  & 200 & 100 & 0.858 & 0.932 & 0.982 & 0.636 & 0.702 & 0.820 \\ 
  & 500 & 100 & 0.900 & 0.956 & 0.982 & 0.514 & 0.580 & 0.678 \\ 
  \hline
  $\nu_N=0$ & 50 & 10 & 0.852 & 0.920 & 0.972 & 0.846 & 0.900 & 0.958 \\ 
  & 100 & 10 & 0.842 & 0.912 & 0.972 & 0.782 & 0.858 & 0.948 \\ 
  & 50 & 50 & 0.858 & 0.916 & 0.982 & 0.808 & 0.882 & 0.948 \\ 
  & 100 & 50 & 0.872 & 0.940 & 0.980 & 0.750 & 0.824 & 0.916 \\ 
  & 200 & 100 & 0.888 & 0.958 & 0.992 & 0.630 & 0.708 & 0.816 \\ 
  & 500 & 100 & 0.880 & 0.948 & 0.990 & 0.494 & 0.556 & 0.656 \\ 
  \hline
  $\nu_N=\mathcal{O}(T^{-1})$ & 50 & 10 & 0.902 & 0.946 & 0.978 & 0.796 & 0.898 & 0.960 \\ 
  & 100 & 10 & 0.880 & 0.932 & 0.982 & 0.812 & 0.886 & 0.950 \\ 
  & 50 & 50 & 0.840 & 0.910 & 0.976 & 0.816 & 0.880 & 0.952 \\ 
  & 100 & 50 & 0.878 & 0.932 & 0.984 & 0.766 & 0.858 & 0.930 \\ 
  & 200 & 100 & 0.902 & 0.944 & 0.988 & 0.750 & 0.838 & 0.938 \\ 
  & 500 & 100 & 0.882 & 0.950 & 0.990 & 0.662 & 0.740 & 0.850 \\ 
 \hline \hline
\end{tabular}
\caption{Empirical coverage of the cross-sectional bootstrap confidence intervals vs. empirical coverage of the asymptotic confidence intervals of \citet{Bai2009} for variable $X_1$.}
\label{table:coverage_compare_strong}
\end{table}

\section{Determinants of Economic Growth}\label{sec5}
The aim of this section is to show the performance of our estimator in empirical analysis. More precisely, we will apply our estimator in the analysis of the determinants of economic growth.  We refer to \citet{durlauf2005growth} for a comprehensive review of the growth literature. While many studies focus on a cross-sectional analysis (see for instance \cite{barro1991economic}), there are also numerous studies employing a panel data approach with country-specific fixed effects \citep{acemoglu2019democracy,islam1995growth}. However, \citet{lu2016shrinkage} argue that economic growth rates might not be solely determined by observable regressors, but could also be influenced by latent factors or shocks. Our projection-based interactive fixed effect estimator is well suited as it is flexible enough to model such latent factors.

\begin{figure}
	\centering
		\includegraphics[width=0.6\textwidth]{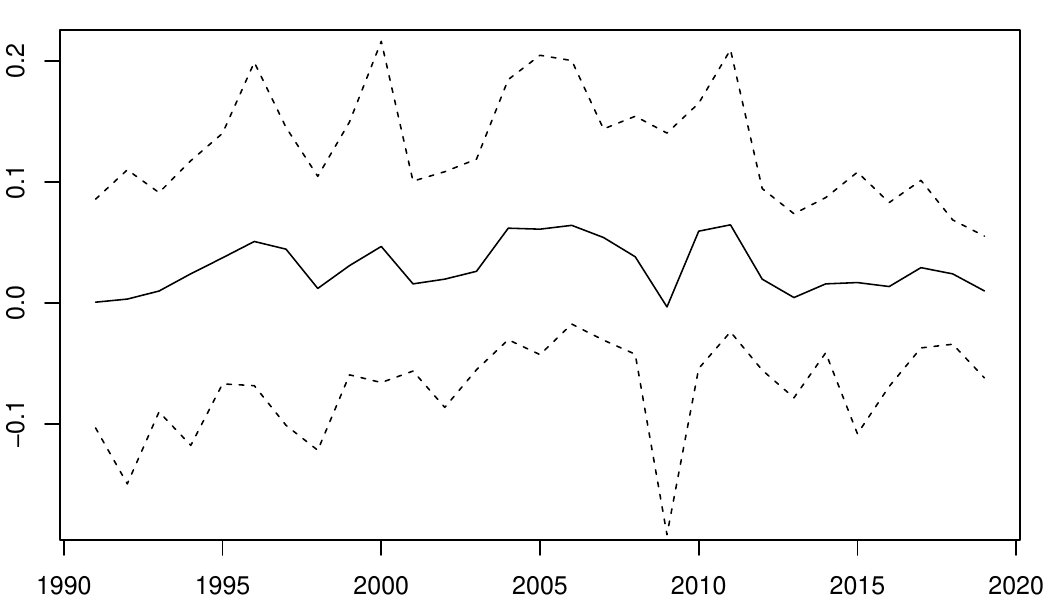}
	\caption{Time series of average annual real GDP growth rate per capita (solid line) and time series of $5\%$ and $95\%$-quantiles (dashed lines).}
	\label{fig:growth}
\end{figure}

The yearly data on GDP growth rates and the country-specific characteristics are obtained from the Penn World Table (PWT) and the World Bank World Development Indicators (WDI). Our sample contains 129 countries in a period from 1991--2019, $N=129$ and $T=29$. Countries with incomplete data availability or which did not exist yet in 1991 are excluded from our analysis. Our dependent variable is the real GDP growth rate per capita. The set of regressors is identical to the regressors in \citet{lu2016shrinkage}. Summary statistics of all dependent and independent variables can be found in Table \ref{table:regressors}. Figure \ref{fig:growth} shows the time series of the mean growth rates, averaged over all countries in our sample. We also visualize the time series of the cross-sectional $5\%$ and $95\%$-quantiles of the growth rates in the same figure. For the time-invariant characteristics used for modeling the systematic part of the factor loadings we take the longitude and latitude of the respective country\footnote{Data obtained from developers.google.com/public-data/docs/canonical/countries\_csv.}.

\begin{table}[tbp]
\footnotesize
\centering
\begin{tabular}{ll|rrrrl}
	\hline \hline
  Variable & Description & Mean & Median & Min & Max & Data \\
  \hline
Growth & Annual GDP growth per capita  &  2.96 & 2.54 & --67.29 & 141.63 & PWT \\
Young & Age dependency ratio &  54.13 & 49.92 & 14.92 & 107.40  & WDI\\
Fert & Fertility rate &  3.23 & 2.69 & 1.09 & 7.7  & WDI \\
Life & Life expectancy &  68.30 & 71.21 & 26.17 & 84.36  &  WDI \\
Pop & Population growth &  1.70 & 1.51 & -6.54 & 19.14 & PWT \\
Invpri & Price level of investment &  0.54 & 0.50 & 0.01 & 7.98 & PWT \\
Con & Consumption share &  0.64 & 0.65 & 0.09 & 1.56  & PWT \\
Gov & Government consumption share &  0.17 & 0.17 & 0.01 & 0.75  & PWT \\
Inv & Investment share &  0.22 & 0.22 & 0.00 & 0.92  & PWT \\
 \hline \hline
\end{tabular}
\caption{Summary statistics and data sources of dependent and independent variables.}
\label{table:regressors}
\end{table}

We first fit our projection-based interactive fixed effects model using the complete sample of $N=129$ countries. To be consistent with the simulation section, we use B-spline basis functions with $J_N=\lceil N^{1/3}1.5\rceil$. The estimation results can be found in Table \ref{table:model_all}. We report the estimated coefficients and the $95\%$ confidence interval based on the cross-sectional bootstrap with $1000$ bootstrap iterations. As a comparison, we also report the estimated coefficients and confidence intervals following the PC-IFE approach of \citet{Bai2009}. We obtain negative significant coefficients for consumption share and government consumption share and a positive significant coefficient for the age dependency ratio at the $5\%$ confidence level. These results are similar to the estimation based on the PC-IFE. However, investment share and fertility rate also become significant for the PC-IFE. The remaining variables are insignificant for both estimation procedures.

\begin{table}[tbp]
\centering
\footnotesize
\begin{tabular}{l|lc|lc}
	\hline\hline
      \multicolumn{1}{r}{} & \multicolumn{2}{c}{P-IFE} & \multicolumn{2}{c}{PC-IFE} \\
	& Estimate & $95\%$-CI & Estimate & $95\%$-CI \\
	\hline
	Con & --0.0508$^{**}$  & [--0.0841, --0.0188] & --0.0518$^{***}$ & [--0.0757, --0.0279] \\
	Gov & --0.0697$^{*}$  & [--0.1374, --0.0125] & --0.1389$^{***}$ & [--0.1830, --0.0948] \\
	Inv & \ 0.0129  & [--0.0420, 0.0592] & \ 0.0503$^{**}$ & [0.0178, 0.0828] \\
	Invpri & \ 0.0076  & [--0.0101, 0.0240 ] & \ 0.0067 & [--0.0042, 0.0176] \\
	Young & \ 0.0007$^{*}$  & [0.0001, 0.0014] & \ 0.0009$^{***}$ & [0.0005, 0.0013] \\
	Fert  & --0.0080  & [--0.0182, 0.0025] & --0.0158$^{***}$ & [--0.0223, --0.0093] \\
	Life & \ 0.0001  & [--0.0008, 0.0010] & \ 0.0000 & [--0.0006, 0.0006] \\
	Pop & --0.0046  & [--0.6075, 0.7627] & --0.2226 & [--0.4696, 0.0244] \\
	\hline\hline
\end{tabular}
\caption{Estimation results for the P-IFE and the PC-IFE based on the whole sample. $^*,^{**},^{***}$ indicate the significance at 5\%, 1\% and 0.1\% level.}
\label{table:model_all}
\end{table}

We now restrict our analysis to the subset of countries that are members of the OECD (Organisation for Economic Cooperation and Development). See Table \ref{table:model_oecd} for the estimation results. Similar to the previous results, both approaches find a negative significant effect on the consumption share. However, our P-IFE identifies a positive effect on investment share and a negative effect on the investment price level. Both variables are insignificant for the PC-IFE approach. Moreover, PC-IFE additionally finds significant effects on the fertility rate and life expectancy.

\begin{table}[tbp]
\centering
\footnotesize
\begin{tabular}{l|lc|lc}
	\hline\hline
      \multicolumn{1}{r}{} & \multicolumn{2}{c}{P-IFE} & \multicolumn{2}{c}{PC-IFE} \\
	& Estimate & $95\%$-CI & Estimate & $95\%$-CI \\
	\hline
	Con & --0.1078$^{***}$  & [--0.1685, --0.0447] & --0.0725$^{***}$ & [--0.1048, --0.0402] \\
	Gov & --0.0343  & [--0.1859, 0.0713] & \ 0.0110 & [--0.0413, 0.0633] \\
	Inv & \ 0.1142$^{*}$ & [0.0117, 0.2226] & \ 0.0405 & [--0.0105, 0.0915] \\
	Invpri & --0.0550$^{*}$  & [--0.0939, --0.0097] & --0.0082 & [--0.0242, 0.0078] \\
	Young & \ 0.0011  & [--0.0001, 0.0021] & \ 0.0008 & [--0.0001, 0.0017] \\
	Fert  & --0.0086 & [--0.0289, 0.0081] & --0.0206$^{*}$ & [--0.0369, --0.0043] \\
	Life & --0.0004 & [--0.0031, 0.0012] & --0.0025$^{***}$ & [--0.0038, --0.0012] \\
	Pop & --0.8195  & [--1.6591, 0.2194] & \ 0.3263 & [--0.1794, 0.8320] \\
	\hline\hline
\end{tabular}
\caption{Estimation results for the P-IFE and the PC-IFE based on OECD sample. $^*,^{**},^{***}$ indicate the significance at 5\%, 1\% and 0.1\% level.}
\label{table:model_oecd}
\end{table}

\section{Conclusions}\label{sec6}
In this paper, a new estimator for the regression parameters in a panel data model with interactive fixed effects has been proposed. The main novelty of this approach is that factor loadings are approximated through nonparametric additive functions, and it is then possible to partial out the interactive effects. Therefore, the new estimator adopts the well-known partial least squares form, and there is no need to use iterative estimation techniques to compute it. It turns out that the limiting distribution of the estimator has a discontinuity when the variance of the idiosyncratic parameter of the factor loading approximation is near the boundaries. The discontinuity makes the usual ``plug-in'' inference based on the estimated asymptotic covariance matrix problematic since it can lead to either over- or under-coveraging probabilities. We show that non-conservative uniformly valid inference can be achieved by cross-sectional bootstrap. A Monte Carlo study indicates good performance in terms of mean squared error and bootstrap coverage. We apply our methodology to analyze the determinants of growth rates in OECD countries.

\section*{Acknowledgements}
Juan M. Rodriguez-Poo and Alexandra Soberon acknowledge financial support from the I+D+i project Ref. PID2019-105986GB-C22 financed by MCIN/AEI/10.13039/ 501100011033. In addition, this work was also funded by the I+D+i project Ref. TED2021-131763A-I00 financed by MCIN/AEI/10.13039/501100011033 and by the European Union NextGenerationUE/PRTR. Georg Keilbar acknowledges gratefully the support from the Deutsche Forschungsgemeinschaft via the IRTG 1792 "High Dimensional Nonstationary Time Series". Weining Wang's research is partially supported by the ESRC (Grant Reference: ES/T01573X/1).








\spacingset{1.3}
\bibliographystyle{ecta}

\bibliography{biball_intro}

{
  \bigskip
  \bigskip
  \bigskip
  \begin{center}
    {\LARGE\bf A projection based approach for interactive fixed effects panel data models}
\end{center}
  \medskip
\centerline{\textbf{SUPPLEMENTARY MATERIAL}}
  \medskip
}

\spacingset{1.8} 

This supplementary document is organized as follows. In Section \ref{appendix_A} we provide the proof of lemmas. The proofs for Theorems 3.1 and 3.2 are in Section \ref{appendix_B}. We provide more details on the estimation of latent factors and factor loadings in Section \ref{appendix_C}. Finally, we provide more details on the simulation study in Section \ref{appendix_D}.

\appendix
\section{Proofs of Lemmas}\label{appendix_A}
\renewcommand{\theequation}{A.\arabic{equation}}
\renewcommand{\thelemma}{A.\arabic{lemma}}
\setcounter{equation}{0}

Through the Appendix, $C$ denotes a generic positive constant that may be different in different uses and $\mathbf{A}$ is a matrix. Also, when we write $\mathbf{A}\leq C$ for a constant scalar $C$, it means that each element of $\mathbf{A}$ is less or equal to $C$. For scalars or column vectors $A_{it}$ and $B_{it}$, we define $S_{A,B}=(NT)^{-1}\sum_{i,t}\mathbf{A}_{it}\mathbf{B}_{it}^{\top}$ and $S_A=S_{A,A}$. We also define the scalar function $\overline{S}_A=(NT)^{-1}\sum_{i,t}\mathbf{A}_{it}^{\top}\mathbf{A}_{it}$, which is the sum of the diagonal elements of $S_A$. Using $ab\leq(a^2+b^2)/2$ we may write $S_{A,B}\leq \overline{S}_A+\overline{S}_B$, so the scalar bound $\overline{S}_A+\overline{S}_B$ bounds each of the elements in $S_{A,B}$. Therefore, if $\overline{S}_A+\overline{S}_B=\mathcal{O}_p(a_n)$, then each element of $S_{A,B}$ is at most $\mathcal{O}_p(a_n)$, which implies that $S_{A,B}=\mathcal{O}_p(a_n)$, where $a_n$ is a positive sequence depending on $n$. Similarly, using the Cauchy-Schwarz inequality we have $S_{A,B}\leq (\overline{S}_A\overline{S}_B)^{1/2}$. Also, for any matrix $\mathbf{A}$ with $N$ rows, we define $\widetilde{\mathbf{A}}=\mathbf{P}_{\Phi}(\mathbf{Z})\mathbf{A}$.

Further, we denote $\theta(z)=\E[\mathbf{X}_t|\mathbf{Z}=z]$ and $m(z)$ is the projection of $\theta(z)$ onto $\mathcal{G}$, i.e., $m(z)=\E_{\mathcal{G}}[\theta(z)]$. Hence, we define $\boldsymbol{\xi}_t = \mathbf{X}_t -\boldsymbol{m}(\mathbf{Z})$, $\boldsymbol{\eta}(\mathbf{Z}) = \boldsymbol{\theta}(\mathbf{Z})-\boldsymbol{m}(\mathbf{Z})$, and
$\boldsymbol{\varepsilon}_t= \mathbf{X}_t -\boldsymbol{\theta}(\mathbf{Z})$, where $\boldsymbol{\xi}_t$, $\boldsymbol{\eta}(\mathbf{Z})$, and $\boldsymbol{\varepsilon}_t$ are $N\times Q$ matrices.

\begin{lemma}\label{lemA2}
Under Assumptions \ref{asum3}(i) and \ref{asum4}(ii), we have,
$\|\mathbf{\widetilde{b}}_h-\mathbf{b}_h\|_F=\mathcal{O}_p(J_N^{-\kappa/2})$, where $\mathbf{\widetilde{b}}_h=\left[\Phi(\mathbf{Z})^{\top}\Phi(\mathbf{Z})\right]^{-1}
\Phi(\mathbf{Z})^{\top}\mathbf{h}$ for $\mathbf{h}=\mathbf{G}(\mathbf{Z})$ or $\mathbf{h}=\boldsymbol{m}(\mathbf{Z})$, and $\mathbf{b}_h$ is the corresponding $J_ND\times K$ or $J_ND\times D$ matrix of sieve coefficients, as $N$ tends to infinity.
\end{lemma}

\begin{proof}
By Assumptions \ref{asum3}(i) and \ref{asum4}(ii)
\begin{eqnarray*}
\|\mathbf{\widetilde{b}}_h-\mathbf{b}_h\|_F &=&\left\|\left[\Phi(\mathbf{Z})^{\top}\Phi(\mathbf{Z})\right]^{-1}
\Phi(\mathbf{Z})^{\top}[\mathbf{h}-\Phi(\mathbf{Z})\mathbf{b}_h]\right\|_F \\
&\leq &\left\|\left[\Phi(\mathbf{Z})^{\top}\Phi(\mathbf{Z})\right]^{-1}
\right\|_2\|\Phi(\mathbf{Z})\|_2\|[\mathbf{h}-\Phi(\mathbf{Z})
\mathbf{b}_h]\|_F \\
&=&\mathcal{O}_p(J_N^{-\kappa/2}).
\end{eqnarray*}
\end{proof}

\begin{lemma}\label{lemA3}
Under Assumptions \ref{asum3}(i) and \ref{asum4}(ii), $\bar{S}_{h-\widetilde{h}} = N^{-1} \|\mathbf{h}-\mathbf{\widetilde{h}}\|^2_F = \mathcal{O}_p(J_N^{-\kappa})$, as $N$ tends to infinity,
for $\mathbf{h}=\mathbf{G}(\mathbf{Z})$ or $\mathbf{h}=\boldsymbol{m}(\mathbf{Z})$, where $\mathbf{\widetilde{h}}=\Phi(\mathbf{Z})\mathbf{\widetilde{b}}_h$.
\end{lemma}

\begin{proof}
Let
\begin{eqnarray*}
\|\mathbf{h}-\mathbf{\widetilde{h}}\|_F & = &
\left\|\left[\mathbf{h}-\Phi(\mathbf{Z})\mathbf{b}_h\right]+
\Phi(\mathbf{Z})(\mathbf{b}_h-\mathbf{\widetilde{b}}_h)\right\|_F \\
& \le & \|\mathbf{h}-\Phi(\mathbf{Z})\mathbf{b}_h\|_F + \|\Phi(\mathbf{Z})(\mathbf{b}_h-\mathbf{\widetilde{b}}_h) \|_F.
\end{eqnarray*}
By Lemma \ref{lemA2} and Assumption \ref{asum3}(i), $\|\Phi(\mathbf{Z})(\mathbf{b}_h-\mathbf{\widetilde{b}}_h)\|_F = \mathcal{O}_p(\sqrt{N}J_N^{-\kappa/2})$ and by Assumption \ref{asum4}(ii), $\|\mathbf{h}-\Phi(\mathbf{Z})\mathbf{b}_h\|_F = \mathcal{O}_p(\sqrt{N}J_N^{-\kappa/2})$, as $N$ tends to infinity.
\end{proof}

\begin{lemma}\label{lemA4}
Under Assumptions \ref{asum3}(i), \ref{asum3}(ii),  \ref{asum6}(i) and \ref{asum6}(iv), $\overline{S}_{\widetilde{U}}=O_p\left(J_N/N\right)$, as both $N$ and $T$ tend to infinity,
where $\overline{S}_{\widetilde{U}}=(NT)^{-1}\sum_{t=1}^T\|\mathbf{\widetilde{u}}_t\|^2$, and $\mathbf{\widetilde{u}}_{t}=\mathbf{P}_{\Phi}(\mathbf{Z})\mathbf u_t$.
\end{lemma}

\begin{proof}
Let
\begin{eqnarray*}
\overline{S}_{\widetilde{U}} & = & (NT)^{-1}\sum_{t=1}^T\frac{\mathbf u_t^{\top}\Phi(\mathbf{Z})}{\sqrt{N}}
\left[\frac{\Phi(\mathbf{Z})^{\top}\Phi(\mathbf{Z})}{N}\right]^{-1}\frac{\Phi(\mathbf{Z})^{\top}\mathbf u_t}{\sqrt{N}}  \\
& = & (NT)^{-1}\sum_{t=1}^T \Big\|\left[\frac{\Phi(\mathbf{Z})^{\top}\Phi(\mathbf{Z})}{N}\right]^{-1/2}\frac{\Phi(\mathbf{Z})^{\top}\mathbf u_t}{\sqrt{N}}\Big\|^2 \\
&\leq& (NT)^{-1}\sum_t\Big\|\left[\frac{\Phi(\mathbf{Z})^{\top}\Phi(\mathbf{Z})}{N}\right]^{-1/2}
\Big\|^2_2\Big\|\frac{\Phi(\mathbf{Z})^{\top}\mathbf u_t}{\sqrt{N}}\Big\|^2.
\end{eqnarray*}

By Assumption \ref{asum3}(i), $\Big\|\left[\Phi(\mathbf{Z})^{\top}\Phi(\mathbf{Z})/N\right]^{-1/2}\Big\|^2_2 = \CO_p\left(1\right)$, as $N$ tends to infinity and
\begin{eqnarray*}
\E\Big\|\frac{\Phi(\mathbf{Z})^{\top}\mathbf u_t}{\sqrt{N}}\Big\|^2 & = & N^{-1}\sum^{J_N}_j\sum^D_d \E\left[\sum^N_{i=1} \phi_j(Z_{id})u_{it}\right]^2 \\
& = & N^{-1}\sum_j\sum_d \sum^N_{i=1} \sum^N_{i^{\prime}=1} \E \left[\phi_j(Z_{id})\phi_j(Z_{i^{\prime}d})\right] \E\left(u_{it}u_{i^{\prime}t}\right)  \\
&\le & J_ND \max_{j,d,i} \E[\phi_j(Z_{id})^2] \max_{i^{\prime}} \sum^N_{i=1} \Big|\E(u_{it} u_{i^{\prime}t})\Big| = \CO\left(J_N\right).
\end{eqnarray*}
Because of Assumptions \ref{asum3}(ii), \ref{asum6}(i) and \ref{asum6}(iv). Then, to Markov's inequality
\[
\Big\|\frac{\Phi(\mathbf{Z})^{\top}\mathbf u_t}{\sqrt{N}}\Big\|^2 = \CO_p\left(J_N\right),
\]
and the proof is done.
\end{proof}

\begin{lemma}\label{lemA5}
Under assumptions \ref{asum0}(i), \ref{asum3}(i) and \ref{asum3}(ii), $\bar S_{\widetilde{\eta}} = N^{-1}\|\boldsymbol{\widetilde{\eta}}\|^2_F =  \mathcal{O}_p\left(J_N/N\right)$, as $N$ tends to infinity, where
$\boldsymbol{\eta}=(\boldsymbol\eta_1,\ldots,\boldsymbol\eta_N)^{\top}$ is an $N\times Q$ matrix, $\boldsymbol{\eta}_i=\boldsymbol\theta(\mathbf{Z}_i)-\boldsymbol{m}(\mathbf{Z}_i)$ and $\boldsymbol{\widetilde{\eta}}=\mathbf{P}_{\Phi}(\mathbf{Z})\boldsymbol{\eta}$.
\end{lemma}

\begin{proof}
The proof follows the same lines as in the proof of Lemma \ref{lemA4}. Therefore we omit the common parts. Noting that
\[
\bar S_{\widetilde{\eta}} = N^{-1}\left\|\left[\frac{\Phi(\mathbf{Z})^{\top}\Phi(\mathbf{Z})}
{N}\right]^{-1/2}\frac{\Phi(\mathbf{Z})^{\top}\boldsymbol\eta}{\sqrt{N}}
\right\|^2_F,
\]
by Assumption \ref{asum3}(i), $\Big\|\left[\Phi(\mathbf{Z})^{\top}\Phi(\mathbf{Z})/N\right]^{-1/2}\Big\|^2_2 = \CO_p\left(1\right)$, as $N$ tends to infinity.  Under assumptions  \ref{asum0}(i) and \ref{asum3}(ii) we obtain that
\begin{eqnarray*}
\E\left\|\frac{\Phi(\mathbf{Z})^{\top}\boldsymbol\eta}{\sqrt{N}}\right\|_F^2 &=& \frac{1}{N}{\rm tr}\E[\Phi(\mathbf{Z})\Phi(\mathbf{Z})^{\top}\boldsymbol\eta
\boldsymbol\eta^{\top}]
\leq \frac{C}{N}{\rm tr}\E[\Phi(\mathbf{Z})\Phi(\mathbf{Z})^{\top}] = \mathcal{O}(J_N),
\end{eqnarray*}
and applying Markov's inequality we have $\Big\|\Phi(\mathbf{Z})^{\top}\boldsymbol\eta/\sqrt{N}\Big\|^2_F=\mathcal{O}_p(J_N)$ and the proof of the lemma is done.
\end{proof}

\begin{lemma}\label{lemA6}
Under assumptions \ref{asum0}(ii), \ref{asum3}(i) and \ref{asum3}(ii), $\bar S_{\widetilde{\varepsilon}}= (NT)^{-1}\sum^T_{t=1}\|\boldsymbol{\widetilde{\varepsilon}}_t\|^2_F =  \mathcal{O}_p\left(J_N/N\right)$, as both $N$ and $T$ tend to infinity, where $\boldsymbol{\varepsilon}=(\boldsymbol\varepsilon_{1t},\ldots,\boldsymbol\varepsilon_{Nt})^{\top}$ is a $N\times Q$ matrix, $\boldsymbol{\varepsilon}_{it}=\mathbf{X}_{it}-\theta(\mathbf{Z}_i)$ and $\boldsymbol{\widetilde{\varepsilon}}_t=\mathbf{P}_{\Phi}(\mathbf{Z})
\boldsymbol\varepsilon_t$.
\end{lemma}

\begin{proof}
Let
\begin{eqnarray*}
\bar S_{\widetilde{\varepsilon}}& = & (NT)^{-1}\sum_{t=1}^T\Big\|\left[\Phi(\mathbf{Z})^{\top}
\Phi(\mathbf{Z})/N\right]^{-1/2}\Phi(\mathbf{Z})^{\top}
\boldsymbol\varepsilon_t/\sqrt{N}\Big\|^2_F.
\end{eqnarray*}

Following a similar reasoning as in Lemmas \ref{lemA4} and \ref{lemA5} it is straightforward to show that under assumptions  \ref{asum0}(ii) and \ref{asum3}(ii), $\|\Phi(\mathbf{Z})^{\top}\boldsymbol\varepsilon_t/\sqrt{N}\|_F^2=
\CO_p\left(J_N\right)$, as $N\rightarrow\infty$. Then the proof is closed.
\end{proof}

\begin{lemma}\label{lemA7}
Under Assumptions  \ref{asum2}, \ref{asum3}(ii),  \ref{asum5}(i) and \ref{asum5}(ii),
\[
\bar{S}_{\widetilde{\Gamma}f} = (NT)^{-1}\sum_{t=1}^T\|\widetilde{\boldsymbol \Gamma}\mathbf{f}_t\|^2 = \CO_p\left(J_N\nu_N/N\right),
\]
as both $N$ and $T$ tend to infinity and $\widetilde{\boldsymbol \Gamma}=\mathbf{P}_{\Phi}(\mathbf{Z})\boldsymbol \Gamma$.
\end{lemma}

\begin{proof}
Proceeding as in the proof of previous lemmas, by idempotence of $\mathbf{P}_{\Phi}(\mathbf{Z})$ we have
\begin{eqnarray*}
\overline{S}_{\widetilde{\Gamma}f}  & = & (NT)^{-1}\sum_{t=1}^T\Big\|\left[\Phi(\mathbf{Z})^{\top}\Phi(\mathbf{Z})/N\right]^{-1/2}\Phi(\mathbf{Z})^{\top}\boldsymbol \Gamma \mathbf{f}_t/\sqrt{N}\Big\|^2 \\
& \le & (NT)^{-1}\sum_{t=1}^T\Big\|\left[\Phi(\mathbf{Z})^{\top}\Phi(\mathbf{Z})/N\right]^{-1/2}\Big\|^2_2\Big\|\Phi(\mathbf{Z})^{\top}\boldsymbol \Gamma \mathbf{f}_t/\sqrt{N}\Big\|^2_F \\
& = & N^{-1}\Big\|\left[\Phi(\mathbf{Z})^{\top}\Phi(\mathbf{Z})/N\right]^{-1/2}\Big\|^2_2\Big\|\Phi(\mathbf{Z})^{\top}\boldsymbol \Gamma/\sqrt{N}\Big\|^2_F.
\end{eqnarray*}
The last equality is obtained by using Assumption \ref{asum2} and applying the properties of the trace operator.

Now, by Assumption \ref{asum3}(i), $\Big\|\left[\Phi(\mathbf{Z})^{\top}\Phi(\mathbf{Z})/N\right]^{-1/2}\Big\|^2_2 = \CO_p\left(1\right)$, as $N$ tends to infinity, and
since $Z_{id}$ and $\gamma_{ik}$ are independent and $\E(\gamma_{ik})=0$, (see Assumption (\ref{asum5})(i) and applying Assumptions \ref{asum3}(ii) and \ref{asum5}(ii)
\begin{eqnarray*}
\E\Big\|\Phi(\mathbf{Z})^{\top}\boldsymbol \Gamma\Big\|^2_F & = & \sum_{j=1}^{J_N}\sum_{k=1}^K\sum_{d=1}^D\E\left[\sum^N_{i=1}\phi_j(Z_{id}) \gamma_{ik}\right]^2 =  \sum_{j=1}^{J_N}\sum_{k=1}^K\sum_{d=1}^D
{\sf Var}\left[\sum^N_{i=1}\phi_j(Z_{id}) \gamma_{ik}\right] \\
& = & \sum_{j=1}^{J_N}\sum_{k=1}^K\sum_{d=1}^D\sum^N_{i=1}{\sf Var}\left[\phi_j(Z_{id})\gamma_{ik}\right] + \sum_{j=1}^{J_N}\sum_{k=1}^K\sum_{d=1}^D\sum^N_{i\ne i^{\prime}}{\sf Cov}\left[\phi_j(Z_{id})\gamma_{ik},\phi_j(Z_{i^{\prime}d}) \gamma_{i^{\prime}k}\right] \\
& = & \sum_{j=1}^{J_N}\sum_{k=1}^K\sum_{d=1}^D\sum^N_{i=1}\E[\phi_j(Z_{id})^2]\E(\gamma^2_{ik}) + \sum_{j=1}^{J_N}\sum_{k=1}^K\sum_{d=1}^D\sum^N_{i\ne i^{\prime}}\E\left[\phi_j(Z_{id})\phi_j(Z_{i^{\prime}d})\right]\E\left(\gamma_{ik}\gamma_{i^{\prime}k}\right) \\
& = & \CO\left(NJ_N\nu_N\right) + \CO\left(NJ_N\right)\max_k \max_{i^{\prime}} \sum^N_{i=1} \left|\E(\gamma_{ik}\gamma_{i^{\prime}k})\right|,
\end{eqnarray*}
and $\max_k \max_{i^{\prime}} \sum^N_{i=1} \left|\E(\gamma_{ik}\gamma_{i^{\prime}k})\right| = \CO\left(\nu_N\right)$ because of Assumption \ref{asum5}(ii). Now apply Markov's inequality and the proof is closed.
\end{proof}

\begin{lemma}[Lemma on Berry Esseen Theorem] [Sunklodas (1984)] \label{berryeseen}
Assume that $\left\{ X_{k}\right\}_{k=1,\ldots, n} $ are $\alpha$-mixing
with geometric rate, and that $B_{n}^{2}\geq cn$ for some $c>0$.
The mixing coefficient $\alpha(t)\leq K \exp(-\lambda t)$ for some positive constants $K$ and $\lambda$. Let $\E|X_{k}|^{p}$ for an integer $p>2$.
$B_{n}^{2}=\E(S_{n}^{2})$ and $S_{n}=\sum_{i}X_{k}^{2}$.  Let
$\Delta_{n}(x)=\left|\P\left(S_{n}\leq xB_{n}\right)-\Phi(x)\right|$.
Then we have,
$$
\sup _{-\infty<x<\infty} \Delta_n(x)=\CO\left(\max _{1 \leq k \leq n} \E\left|X_k\right|^p\left(\log B_n\right)^{p-1} B_n^{2-p}\right).
$$
\end{lemma}

\section{Proofs for Section \ref{sec3}}\label{appendix_B}
\renewcommand{\theequation}{I.\arabic{equation}}
\subsection{Proof of Theorem \ref{theo1}}\label{appendix_B1}
Plugging the matrix form of (\ref{eq4}) into (\ref{eq10}) and using the idempotence property of $\mathbf{P}_{\Phi}(\mathbf{Z})$, after rearranging terms we get
\begin{eqnarray}\label{eqA3}
\nonumber \widehat{\boldsymbol{\beta}}-\boldsymbol{\beta} &=&\left[\frac{1}{NT}\sum^T_{t=1}(\mathbf{X}_t-\widetilde{\mathbf{X}}_t)^{\top}
(\mathbf{X}_t-\widetilde{\mathbf{X}}_t)\right]^{-1}\\
\nonumber&\times&
\frac{1}{NT}\sum^T_{t=1}(\mathbf{X}_t-\widetilde{\mathbf{X}}_t)^{\top}\left[(\mathbf{G}(\mathbf{Z})-\widetilde{\mathbf G}(\mathbf{Z}))\mathbf{f}_t + \mathbf{v}_t\right], \\ \hfill \nonumber
\end{eqnarray}
and defining
\begin{eqnarray}
    S_{X-\widetilde{X}} & \stackrel{\text{def}}{=} & \frac{1}{NT}\sum^T_{t=1}(\mathbf{X}_t-\widetilde{\mathbf{X}}_t)^{\top}
    (\mathbf{X}_t-\widetilde{\mathbf{X}}_t), \\
    S_{X-\widetilde{X},(G-\widetilde{G})f} & \stackrel{\text{def}}{=} & \frac{1}{NT}\sum^T_{t=1}(\mathbf{X}_t-\widetilde{\mathbf{X}}_t)^{\top}
    [\mathbf{G}(\mathbf{Z})-\widetilde{\mathbf{G}}(\mathbf{Z})]\mathbf{f}_t, \\
    S_{X-\widetilde{X},V} & \stackrel{\text{def}}{=} & \frac{1}{NT}\sum^T_{t=1}(\mathbf{X}_t-\widetilde{\mathbf{X}}_t)^{\top}
    \mathbf{v}_t,
\end{eqnarray}
we have that
\begin{equation}
\nonumber \widehat{\boldsymbol{\beta}}-\boldsymbol{\beta} =  S^{-1}_{X-\widetilde{X}} \left(S_{X-\widetilde{X},(G-\widetilde{G})f}+S_{X-\widetilde{X},V}\right),
\end{equation}
where $\boldsymbol{\widetilde{G}}(\mathbf{Z})=\mathbf{P}_{\Phi}(\mathbf{Z})\mathbf{G}(\mathbf{Z})$,  $\mathbf{\widetilde{X}}_t= \mathbf{P}_{\Phi}(\mathbf{Z})\mathbf{X}_t$, $\widetilde{\mathbf{v}}_t= \mathbf{P}_{\Phi}(\mathbf{Z})\mathbf{v}_t$, and $\mathbf{v}_t = \boldsymbol \Gamma \mathbf{f}_t + \mathbf u_t$.

We will obtain the asymptotic distribution of $\widehat{\boldsymbol{\beta}}$, by showing the following results:
\begin{itemize}
\item[(i)] $S_{X-\widetilde{X}}= \widetilde{V}_{\xi} + {\scriptstyle\mathcal{O}}_p(1)$,
\item[(ii)] $S_{X-\widetilde{X},(G-\widetilde{G})f}={\scriptstyle\mathcal{O}}_p\left(
    (NT)^{-1/2}\right)$,
\item[(iii)] For $\vartheta \in [0,1)$, $\sqrt{NT^{\vartheta}}S_{(X-\widetilde{X}),V}\stackrel{\mathcal{L}}{\rightarrow} N(0,\widetilde{V}^{-1}_{\xi}\widetilde{V}_{\Gamma}\widetilde{V}^{-1}_{\xi})$,
for $\vartheta = 1$, $\sqrt{NT}S_{(X-\widetilde{X}),V}\stackrel{\mathcal{L}}{\rightarrow} N(0,\widetilde{V}^{-1}_{\xi}\left(\widetilde{V}_{\Gamma}+\widetilde{V}_{u}\right)\widetilde{V}^{-1}_{\xi})$ and for $\vartheta > 1$, $\sqrt{NT}S_{(X-\widetilde{X}),V}\stackrel{\mathcal{L}}{\rightarrow} N(0,\widetilde{V}^{-1}_{\xi}\widetilde{V}_{u}\widetilde{V}^{-1}_{\xi})$.
\end{itemize}

For the sake of simplicity, we will use the following shorthand notations: $\boldsymbol{\theta}_i = \boldsymbol{\theta}(\mathbf{Z}_i)$, $\boldsymbol{m}_i = \boldsymbol{m}(\mathbf{Z}_i)$. Furthermore, recall that
$\boldsymbol\varepsilon_{it} = \mathbf{X}_{it}-\boldsymbol\theta_i$, $\boldsymbol\eta_i = \boldsymbol\theta_i - \boldsymbol{m}_i$ and $\boldsymbol{\xi}_{it} = \mathbf{X}_{it}- \boldsymbol{m}_i$.  Then, we can write $\mathbf{X}_{it} = \boldsymbol\varepsilon_{it} + \boldsymbol\eta_i + \boldsymbol{m}_i$ and $\widetilde{\mathbf{X}}_{it} = \boldsymbol{\widetilde{\varepsilon}}_{it} + \boldsymbol{\widetilde{\eta}}_i + \boldsymbol{\widetilde{m}}_i$. By subtracting these quantities it yields
\begin{equation}\label{eqA2}
\mathbf{X}_{it} - \widetilde{\mathbf{X}}_{it} = (\boldsymbol\varepsilon_{it} - \boldsymbol{\widetilde{\varepsilon}}_{it}) + (\boldsymbol\eta_i - \boldsymbol{\widetilde{\eta}}_i) + (\boldsymbol{\widetilde{m}}_i - \boldsymbol{m}_i).\quad
\end{equation}

\textbf{Proof of (i)}:  Using (\ref{eqA2}), $S_{X-\widetilde{X}}$ can be decomposed as
\begin{eqnarray}\label{eqA4}
S_{X-\widetilde{X}}=S_{\eta+\varepsilon+(m-\widetilde{m})-\widetilde{\eta}-\widetilde{\varepsilon}}
=S_{\eta+\varepsilon}+S_{(m-\widetilde{m})-\widetilde{\eta}-\widetilde{\varepsilon}}+
2S_{\eta+\varepsilon,(m-\widetilde{m})-\widetilde{\eta}-\widetilde{\varepsilon}},
\end{eqnarray}
where each element has to be considered separately. Focusing on the first element of the right-hand side of (\ref{eqA4}), we get
\begin{eqnarray}\label{eqA5}
S_{\eta+\varepsilon}=(NT)^{-1}\sum^T_{t=1}(\boldsymbol\eta+\boldsymbol
\varepsilon_t)^{\top}(\boldsymbol\eta+\boldsymbol{\varepsilon}_t)= (NT)^{-1}
\sum^T_{t=1}\boldsymbol\xi_{t}^{\top}\boldsymbol\xi_{t},
\end{eqnarray}
where $\boldsymbol\eta=(\boldsymbol\eta_1,\ldots,\boldsymbol\eta_N)^{\top}$, $\boldsymbol\varepsilon_t=(\boldsymbol\varepsilon_{1t},\ldots,\boldsymbol\varepsilon_{Nt})^{\top}$ and $\boldsymbol\xi_t=(\boldsymbol\xi_{1t},\ldots,\boldsymbol\xi_{Nt})$ are $N\times Q$ matrices. 

$S_{(m-\widetilde{m})-\widetilde{\eta}-\widetilde{\varepsilon}} = \mathcal{O}_p(J_N^{-2\kappa})+\mathcal{O}_p\left(J_N/N\right)={\scriptstyle\mathcal{O}}_p(1)$, because $S_{(m-\widetilde{m})-\widetilde{\eta}-\widetilde{\varepsilon}} \leq 2\overline{S}_{m-\widetilde{m}}+4\overline{S}_{\widetilde{\eta}}+
4\overline{S}_{\widetilde{\varepsilon}}$. Then, applying respectively Lemmas \ref{lemA3}, \ref{lemA5} and \ref{lemA6} we have that $\overline{S}_{m-\widetilde{m}}=\mathcal{O}_p(J_N^{-\kappa})$, $\overline{S}_{\widetilde{\eta}} = \overline{S}_{\widetilde{\varepsilon}} = \mathcal{O}_p\left(J_N/N\right)$.  Finally, the third term in (\ref{eqA4}) is ${\scriptstyle\mathcal{O}}_p(1)$ because
given that $(\eta+\varepsilon)$ is orthogonal to the functional space $\mathcal{G}$ and $m-\widetilde{m}$ belongs to $\mathcal{G}$, by the Cauchy-Schwarz inequality we obtain that $S_{\eta+\varepsilon,(m-\widetilde{m})-\widetilde{\eta}-\widetilde{\varepsilon}} \leq \overline{S}^{1/2}_{\eta+\varepsilon}\overline{S}^{1/2}_{(m-\widetilde{m})-\widetilde{\eta}-\widetilde{\varepsilon}}=\mathcal{O}_p(1)\times {\scriptstyle\mathcal{O}}_p(1)={\scriptstyle\mathcal{O}}_p(1)$. Now apply Assumption \ref{asum0}(iii) to (\ref{eqA5}) and this ends the proof of (i).

\textbf{Proof of (ii)}: By the idempotence property of $\mathbf M_{\Phi}(\mathbf{Z}) = \mathbf{I}_N - \mathbf{P}_{\Phi}(\mathbf{Z})$
\begin{eqnarray*}	
S_{X-\widetilde{X},(G-\widetilde{G})f}& = &\frac{1}{NT}\sum_{t=1}^{T}\mathbf{X}_t^{\top} \mathbf M_{\Phi}(\mathbf{Z})\mathbf{G}(\mathbf{Z})\mathbf{f}_t \\
         &=&\frac{1}{NT}\sum_{t=1}^{T}\left(\mathbf{X}_t-\widetilde{\mathbf{X}}_t\right)^{\top}
         \left[\mathbf{G}(\mathbf{Z})-\widetilde{\mathbf{G}}(\mathbf{Z})\right]\mathbf{f}_t.
\end{eqnarray*}

Applying the Cauchy-Schwarz inequality then
\[
S_{X-\widetilde{X},(G-\widetilde{G})f} \le \bar S^{1/2}_{(X-\widetilde X)} \bar S^{1/2}_{(G(Z)-\widetilde{G}(Z))f}.
\]

By (i), $\bar S_{(X-\widetilde X)} = \mathcal{O}_p(1)$ and $\bar S_{(G(Z)-\widetilde{G}(Z))f} = \mathcal{O}_p(J_N^{-\kappa})$. This last result follows from recalling that by the trace properties and Lemma \ref{lemA3}
\[
\bar S_{(G(Z)-\widetilde{G}(Z))f} = N^{-1}{\rm tr}\{[\mathbf{G}(\mathbf{Z})-\widetilde{\mathbf{G}}(\mathbf{Z})]
[\mathbf{G}(\mathbf{Z})-\widetilde{\mathbf{G}}(\mathbf{Z})]^{\top}\}.
\]

Finally, since for $\kappa \geq 4$ and $\varrho \in (\frac{1}{\kappa},\frac{1}{2})$,  $J_N \sim N^{\varrho}$
and $T/N^{\kappa\varrho-1}\rightarrow 0$, then $\sqrt{NT}J_N^{-\kappa/2}\to 0$ and we have $S_{X-\widetilde{X},(G-\widetilde{G})f}={\scriptstyle\mathcal{O}}_p\left((NT)^{-1/2}\right)$, as both $N$ and $T$ tend to infinity.

{\textbf{Proof of (iii)}: Using (\ref{eqA2}) and the definitions therein, we have that 
\begin{eqnarray}\label{eqbb1}
S_{(X-\widetilde{X}),V} & = & S_{\eta+\varepsilon + (m-\widetilde{m}) - \widetilde{\eta} - \widetilde{\varepsilon},V}  \nonumber \\
& = & S_{\xi,V} + S_{(m-\widetilde{m}),V} - S_{\widetilde{\xi},V}.
\end{eqnarray}
Let
\begin{eqnarray*}
S_{(m-\widetilde{m}),V} & = & S_{(m-\widetilde{m}), \Gamma f} +  S_{(m-\widetilde{m}),U}.
\end{eqnarray*}
By the Cauchy-Schwarz inequality
\begin{eqnarray*}
S_{(m-\widetilde{m}), \Gamma f} & \le & \bar{S}^{1/2}_{(m-\widetilde{m})} \bar{S}^{1/2}_{ \Gamma f}.
\end{eqnarray*}
A direct application of Lemma \ref{lemA3} gives $\bar{S}_{(m-\widetilde{m})}=\mathcal{O}_p(J_N^{-\kappa})$. Furthermore, under Assumptions \ref{asum2} and \ref{asum5}(ii), $\bar S_{ \Gamma f} = \mathcal{O}_p(\nu_N)$ and therefore 
$S_{(m-\widetilde{m}), \Gamma f} = \Co_p\left(\sqrt{\nu_N/NT}\right)$, because $\sqrt{NT}J_N^{-\kappa/2}\rightarrow 0$. Proceeding in a similar way, using Assumptions \ref{asum6}(i) and \ref{asum6}(iii), we obtain that $\bar S_{U}=\mathcal{O}_p(1)$ and therefore, $S_{(m-\widetilde{m}),V} =  \Co_p\left(\sqrt{\nu_N/NT}\right)+ \Co_p\left(1/\sqrt{NT}\right)$, also because $\sqrt{NT}J_N^{-\kappa/2}\rightarrow 0$. 

Now, for the third term in (\ref{eqbb1}),  $S_{\widetilde{\xi},V}$, we will show that, 
\begin{equation}\label{eqAA1}
S_{\widetilde{\xi},V} = {\scriptstyle\mathcal{O}}_p(1/\sqrt{NT}),
\end{equation}
for $\kappa\ge4$, $\varrho \in (\frac{1}{\kappa},\frac{1}{2})$ and $J_N \sim N^{\varrho}$, as both $N$ and $T$ tend to infinity. 
In order to show it, note that
\begin{eqnarray}\label{eqA7}  
\nonumber   S_{\widetilde{\xi},V} & = & S_{\widetilde{\eta},V} + S_{\widetilde{\varepsilon},V}  \\
& = &  S_{\widetilde{\eta},U} + S_{\widetilde{\eta},\Gamma f} + S_{\widetilde{\varepsilon},U} + S_{\widetilde{\varepsilon},\Gamma f}. 
\end{eqnarray}

The proof follows by noting that by the idempotence property of $\mathbf{P}_{\Phi}(\mathbf{Z})$ and the Cauchy-Schwarz inequality, $S_{\widetilde{\eta},\widetilde{\Gamma}f}  \le  \bar S^{1/2}_{\widetilde{\eta}}\bar S^{1/2}_{\widetilde{\Gamma}f}$,  and $S_{\widetilde{\varepsilon},\widetilde{\Gamma}f}  \le  \bar S^{1/2}_{\widetilde{\varepsilon}}\bar S^{1/2}_{\widetilde{\Gamma}f}$.
A direct application of Lemmas \ref{lemA5} and \ref{lemA7} gives $S_{\widetilde{\eta},\widetilde{\Gamma}f} = \mathcal{O}_p(J_N\nu^{1/2}_N/N)$,  and by applying Lemmas \ref{lemA6} and \ref{lemA7} we obtain $S_{\widetilde{\varepsilon},\widetilde{\Gamma}f} =  \mathcal{O}_p(J_N\nu^{1/2}_N/N)$. For $S_{\widetilde{\eta},U}$ by assumptions \ref{asum0}(i) and \ref{asum6}(i) note that 
\begin{eqnarray*}
\E\left(\left.\left\|S_{\widetilde\eta,U}\right\|_F^2\right| \mathbf{Z}\right) & = & \frac{1}{N^2T^2} \sum_t\sum_s \E\left(\left.\mathbf{u}^{\top}_t\widetilde{\boldsymbol\eta}\widetilde{\boldsymbol\eta}^{\top}\mathbf{u}_s\right|\mathbf{Z}\right) \\
& = & \frac{1}{N^2T} \mathrm{tr}\left\{\widetilde{\boldsymbol\eta}\widetilde{\boldsymbol\eta}^{\top}\frac{1}{T}\sum_t\sum_s \E\left(\mathbf u_t\mathbf u^{\top}_s\right)\right\}.
\end{eqnarray*}

By Assumptions \ref{asum6}(iii) and \ref{asum6}(iv) applying Davydov's inequality (see \citet{MR1640691}, p. 36) gives 
\[
\frac{1}{T}\sum_t\sum_s \E\left(\mathbf{u}_t\mathbf{u}^{\top}_s\right) \le \mathrm{max}_t \sum_s \E\left(\mathbf{u}_t\mathbf{u}^{\top}_s\right) = \mathcal{O}_p\left(1\right).
\]

Then,
\begin{eqnarray*}
\E\left(\left.\left\|S_{\widetilde\eta,U}\right\|_F^2\right| \mathbf{Z}\right) & \le & \frac{C}{N^2T} \mathrm{tr}\left\{\widetilde{\boldsymbol\eta}\widetilde{\boldsymbol\eta}^{\top}\right\} \\
& = &  \frac{C}{NT} \bar{S}_{\widetilde{\eta}} \\
& = & \mathcal{O}_p\left(J_N/N^2T\right).
\end{eqnarray*}

Finally, for $S_{\widetilde{\varepsilon},U}$ note that, by Assumptions \ref{asum6}(i)-(ii), and applying both Davydov's inequality and the Cauchy-Schwarz inequality we obtain
\begin{eqnarray*}
\E\left(\left.\left\|S_{\widetilde\varepsilon,U}\right\|_F^2\right| \mathbf{Z}\right) & = & \frac{1}{N^2T^2} \sum_t\sum_s \E\left(\left.\mathbf{u}^{\top}_t\widetilde{\boldsymbol\varepsilon}_t\widetilde{\boldsymbol\varepsilon}^{\top}_s\mathbf{u}_s\right|\mathbf{Z}\right) \\
& = & \frac{1}{N^2T^2} \sum_t\sum_s\mathrm{tr}\left\{\E\left(\left.\widetilde{\boldsymbol\varepsilon_t}\widetilde{\boldsymbol\varepsilon_s}^{\top} \mathbf u_s\mathbf u^{\top}_t\right|\mathbf{Z}\right)\right\}. \\
& = & \frac{1}{N^2T^2} \sum_t\sum_s\mathrm{tr}\left\{\E\left(\left.\widetilde{\boldsymbol\varepsilon_t}\widetilde{\boldsymbol\varepsilon_s}^{\top}\right| \mathbf{Z}\right)\E\left(\mathbf u_s\mathbf u^{\top}_t\right)\right\} \\
& = & \frac{1}{N^2T} \mathrm{tr}\left\{\E\left(\left.\widetilde{\boldsymbol\varepsilon_t}\widetilde{\boldsymbol\varepsilon_s}^{\top}\right| \mathbf{Z}\right)\frac{1}{T}\sum_t\sum_s\E\left(\mathbf u_s\mathbf u^{\top}_t\right)\right\} \\
& \le &  \frac{C}{N^2T} \mathrm{tr}\left\{\E\left(\left.\widetilde{\boldsymbol\varepsilon_t}\widetilde{\boldsymbol\varepsilon_s}^{\top}\right| \mathbf{Z}\right)\right\} \\
& \le &  \frac{C}{NT} \E\left(\left.\frac{1}{N}\mathrm{tr}\left\{\widetilde{\boldsymbol\varepsilon_t}^{\top}\widetilde{\boldsymbol\varepsilon_t}\right\}\right| \mathbf{Z}\right) \\
& = & \frac{C}{NT} \E\left(\left. \bar{S}_{\widetilde{\varepsilon}} \right| \mathbf{Z}\right),
\end{eqnarray*}
for some $C>0$. Now a direct application of Lemma \ref{lemA6} proofs that $S_{\widetilde{\varepsilon},U} = \mathcal{O}_p\left(J^{1/2}_N/N\sqrt{T}\right)$. By substituting these asymptotic bounds in (\ref{eqA7}) we obtain (\ref{eqAA1}).

Finally we find a convergence in distribution result for $S_{\xi,V}$. In order to do it, let
\begin{eqnarray*}
S_{\xi,V} & = & \frac{1}{NT} \sum_t \boldsymbol{\xi}^{\top}_t\left(\boldsymbol \Gamma \mathbf{f}_t + \mathbf u_t\right) \\
 & = & \frac{1}{NT} \sum_t [\boldsymbol{\xi}^{\top}_t\boldsymbol \Gamma \mathbf{f}_t - \E(\boldsymbol\xi^{\top}_{t}\boldsymbol \Gamma \boldsymbol f_{t} |\boldsymbol \Gamma)] + \frac{1}{NT} \sum_t \boldsymbol{\xi}^{\top}_t \mathbf u_t + \frac{1}{NT}\sum_t \E(\boldsymbol\xi^{\top}_{t}\boldsymbol \Gamma \boldsymbol f_{t} |\boldsymbol \Gamma).
\end{eqnarray*}

Then, note that under Assumptions \ref{asum5}, \ref{asum6}, and \ref{asum66}, to show a limit distribution for $S_{\xi,V} $, it is enough to show that, as both $N$ and $T$ tend to infinity, 
\begin{eqnarray}\label{eqbb4}
\frac{1}{NT} \sum_t [\boldsymbol{\xi}^{\top}_t\boldsymbol \Gamma \mathbf{f}_t - \E(\boldsymbol\xi^{\top}_{t}\boldsymbol \Gamma \mathbf{f}_{t}|\boldsymbol \Gamma)] & = & \mathcal{O}_p\left(\sqrt{\frac{\nu_N}{NT}}\right) \label{eqbb2}, \\
(NT)^{-1/2}\widetilde{V}^{-1/2}_{u} \sum_t \boldsymbol\xi^{\top}_t \mathbf u_t  & \stackrel{\mathcal{L}}{\rightarrow} &  N\left(0,I_Q\right), \label{eqbb3} \\
N^{-1/2}\widetilde{V}^{-1/2}_{\Gamma}T^{-1}  \sum_t \E(\boldsymbol\xi^{\top}_{t}\boldsymbol \Gamma \mathbf{f}_{t}|\boldsymbol \Gamma) & \stackrel{\mathcal{L}}{\rightarrow} &  N\left(0,I_Q\right) \label{eqbb33}.
\end{eqnarray}

To prove (\ref{eqbb2}), we look at the variance of the $q$-th element of the left hand side,
\begin{align*}
    &{\rm Var}\left(\frac{1}{NT}\sum_{t=1}^T\boldsymbol\xi^{\top}_{tq}\Gamma f_t-\E\left(\boldsymbol\xi^{\top}_{tq}\Gamma f_t|\Gamma\right)\right)\\
    &\leq\frac{1}{N^2T^2}\sum_{i=1}^N\sum_{t=1}^T K \max_k\E\left(\boldsymbol\xi_{itq}\gamma_{ik} f_{kt}-\E\left(\boldsymbol\xi_{itq}\gamma_{ik} f_{tk}|\Gamma\right)\right)^2\\
    &+ \frac{1}{N^2T^2}\sum_{i=1}^N\sum_{t=1}^T K\sum_{s\neq t}^T\max_k\E\left(\left[\boldsymbol\xi_{itq}\gamma_{ik} f_{kt}-\E\left(\boldsymbol\xi_{itq}\gamma_{ik} f_{tk}|\Gamma\right)\right]\left[\boldsymbol\xi_{isq}\gamma_{ik} f_{ks}-\E\left(\boldsymbol\xi_{isq}\gamma_{ik} f_{sk}|\Gamma\right)\right]\right)\\
    &\leq\frac{1}{N^2T^2}\sum_{i=1}^N\sum_{t=1}^T\nu_N K \max_k\E\left(\boldsymbol\xi_{itq} f_{kt}-\E\left(\boldsymbol\xi_{itq} f_{tk}|\Gamma\right)\right)^2\\
    &+ \frac{1}{N^2T^2}\sum_{i=1}^N\sum_{t=1}^T\sum_{s\neq t}^T\nu_N K\max_k\E\left(\left[\boldsymbol\xi_{itq} f_{kt}-\E\left(\boldsymbol\xi_{itq} f_{tk}|\Gamma\right)\right]\left[\boldsymbol\xi_{isq}f_{ks}-\E\left(\boldsymbol\xi_{isq} f_{sk}|\Gamma\right)\right]\right)\\
    &\leq\frac{\nu_N}{N^2T}NCM_{\delta}\sum_{-\infty<k<\infty}\left(ak^{-\tau}\right)^{1-2/\delta}\\
    &=\mathcal{O}\left(\frac{\nu_N}{NT}\right).
\end{align*}
The last inequality follows from Assumption 3.5 (i)-(ii), 3.6 (i) and Davydov's inquality, see Theorem 0.1 of \citet{rio1993covariance}. The series $\sum_{-\infty<k<\infty}\left(ak^{-\tau}\right)^{1-2/\delta}$ is absolutely convergent by Assumption 3.6 (iii).

In order to show (\ref{eqbb3}), we employ the Cram\'{e}r-Wold device because $(NT)^{-1/2}\sum_t \boldsymbol\xi^{\top}_t\mathbf u_t$ is multivariate. For any unit vector $\mathbf{e} \in \mathbb{R}^Q$, let
$\omega_t = N^{-1/2}\sum_i\mathbf{e}^{\top}\boldsymbol\xi_{it}u_{it}$ and then
\[
\frac{1}{\sqrt{NT}} \sum^T_{t=1} \mathbf{e}^{\top}\boldsymbol\xi^{\top}_t\mathbf{u}_t = \frac{1}{\sqrt{T}} \sum^T_{t=1} \omega_t.
\]

Note that by Assumption \ref{asum6}(i),  $\left\{\boldsymbol\xi_{it}u_{it}\right\}^T_{t=1}$ is a $\mathbb{R}^Q$-stationary $\alpha$-mixing sequence. Furthermore, by Assumptions \ref{asum6}(ii) and \ref{asum66}(i), a direct application of Theorem 1.5 in \citet{MR1640691}, p. 34, gives
\[
{\sf Var}\left(\frac{1}{\sqrt{T}} \sum^T_{t=1} \omega_t\right) = \mathbf{e}^{\top}\widetilde{V}_u \mathbf{e} + {\scriptstyle\mathcal{O}}_p(1),
\]
because by Assumption \ref{asum6}(iii), $\sum_{k\ge 1}\alpha(k)^{1-\frac{1}{\delta}} < \infty$.

Now we prove the asymptotic normality of $T^{-1/2}\sum^T_{t=1} \omega_t$. Considering Assumptions \ref{asum6}(i)-(iii), \ref{asum66}(i) and that $\mathbf{e}^{\top}\widetilde{V}_u\mathbf{e} < \infty$ a direct application of a Central Limit Theorem for strongly mixing processes (see Theorem 1.7 in \citet{MR1640691}, p. 36) gives
\begin{eqnarray}
   \frac{1}{\sqrt{T}}\sum^T_{t=1} \omega_t & \stackrel{\mathcal{L}}{\rightarrow}  & N\left(0,\mathbf{e}^{\top}\widetilde{V}_u\mathbf{e}\right),
\end{eqnarray}
as both $N$ and $T$ tend to infinity.

To obtain the asymptotic distribution of $(NT)^{-1/2}\sum^T_{t=1} \E\left(\left.\boldsymbol\xi^{\top}_{t}\boldsymbol \Gamma \mathbf{f}_{t} \right|\boldsymbol \Gamma\right)$ note that it is multivariate and therefore we need to use again the Cram\'{e}r-Wold device. For any unit vector $\mathbf{e} \in \mathbb{R}^Q$, let
\begin{eqnarray*}
\frac{1}{NT}\sum^T_{t=1} \E\left(\left.\mathbf{e}^{\top}\boldsymbol\xi^{\top}_{t}\boldsymbol \Gamma \mathbf{f}_{t} \right|\boldsymbol \Gamma\right) & = & \frac{1}{N}\sum^N_{i=1}\sum^K_{k=1} \gamma_{ik}\frac{1}{T}\sum^T_{t=1} \E\left(\left.\mathbf{e}^{\top}\boldsymbol\xi_{it}f_{tk}\right| \boldsymbol \Gamma\right) \\
& = & \frac{1}{N}\sum^N_{i=1}\left(\sum^K_{k=1} \gamma_{ik}C_{T,ik}\right) \\
& = &  \frac{1}{N}\sum^N_{i=1}\rho_i.
\end{eqnarray*}

In this case, under Assumption \ref{asum66}, $\rho_1,\ldots,\rho_N$ are independent and non-identically distributed $\mathbb{R}^Q$-random variables. In order to show (\ref{eqbb33}) we rely on a corollary to Liapounov's Central Limit Theorem for independent and heterogeneously distributed observations (see \citet{loeve77}, p. 287). 

Note that, we have that $\E\left(\rho_i\right) = 0$. In order to show it, let
\begin{eqnarray*}
\E\left(\rho_i\right) & = & \frac{1}{T}\sum^K_{k=1} \sum^T_{t=1} \E\left[\gamma_{ik}\E\left(\left.\mathbf{e}^{\top}\boldsymbol\xi_{it}f_{tk}\right| \boldsymbol \Gamma\right)\right]  \\
& = & \frac{1}{T}\sum^K_{k=1} \sum^T_{t=1} \E\left[f_{tk}\E\left(\left.\mathbf{e}^{\top}\boldsymbol\xi_{it}\gamma_{ik}\right| f_{tk}\right)\right]  \\
& = & \frac{1}{T}\sum^K_{k=1} \sum^T_{t=1} \E\left[f_{tk}\E\left(\left.\mathbf{e}^{\top}\boldsymbol\xi_{it}\right|f_{tk}\right)\E\left(\left.\gamma_{ik}\right| f_{tk}\right)\right]  \\
& = & 0.
\end{eqnarray*}
The last equality holds because of Assumption \ref{asum5}(i). Furthermore, 
\begin{equation}
    {\sf Var}\left(\rho_i\right) = \sum_{k\ne k^{\prime}}\E\left[\gamma_{ik}\gamma_{ik^{\prime}}\frac{1}{T^2}\sum_{t,s}\E\left(\left.\mathbf{e}^{\top}\boldsymbol\xi_{it}f_{tk}\right| \boldsymbol \Gamma\right)\E\left(\left.\mathbf{e}^{\top}\boldsymbol\xi_{is}f_{sk^{\prime}}\right| \boldsymbol \Gamma\right)\right] > 0
\end{equation}
and for some  $\delta > 2$, and for all $i$,
\begin{eqnarray*}
\E\left|\rho_i\right|^{\delta} & = & \E\left|\sum_{k} \gamma_{ik}T^{-1}\sum_t  \E\left(\left.\mathbf{e}^{\top}\boldsymbol\xi_{it}f_{tk}\right| \boldsymbol \Gamma\right)\right|^{\delta} \le \sum_k \E\left| \gamma_{ik}\frac{1}{T}\sum_t\E\left(\left.\mathbf{e}^{\top}\boldsymbol\xi_{it}f_{tk}\right| \boldsymbol \Gamma\right)\right|^{\delta} \\
& \le &K\max_{k\le K} \E\left| \gamma_{ik}\frac{1}{T}\sum_t\E\left(\left.\mathbf{e}^{\top}\boldsymbol\xi_{it}f_{tk}\right| \boldsymbol \Gamma\right)\right|^{\delta} < \infty,
\end{eqnarray*}
by assumption \ref{asum5}(iii). To show that the CLT applies we need to check that, for some $\delta>0$, then 
\[
\frac{1}{N}\sum^N_{i=1}\sf Var\left(\rho_i\right) > \delta > 0,
\]
for all $N$ sufficiently large. But note that, under assumptions \ref{asum5}(ii), \ref{asum6}(i) and \ref{asum6}(iii) the expression above holds. Then the CLT applies and
\begin{equation*}
N^{-1/2} \sum^N_{i=1}\rho_i \stackrel{\mathcal{L}}{\rightarrow} N\left(0,\mathbf{e}^{\top}\widetilde{V}_{ \Gamma}\mathbf{e}\right),
\end{equation*}
as both $N$ and $T$ tend to infinity.

To obtain $\widetilde{V}_{\Gamma}$ note that by Assumptions \ref{asum6}(i) and \ref{asum66}(ii)
\begin{eqnarray*}
\widetilde{V}_{\Gamma} = \lim_{N \rightarrow \infty}\left(N^{-1} \sum^N_{i=1}{\sf Var}\left(\rho_i\right)\right) & = &  \lim_{N \rightarrow \infty}\frac{1}{N}\sum^N_{i=1}\E\left[\left(\sum^K_{k=1} \gamma_{ik}C_{T,ik}\right)\left(\sum^K_{k=1} \gamma_{ik}C_{T,ik}\right)^{\top}\right] \\
& = & \lim_{N \rightarrow \infty}\frac{1}{N}\sum_{k\le K; k^{\prime} \le K} \E\left[ \gamma_{ik} \gamma_{ik^{\prime}}\frac{1}{T^2}\sum_{t,s}\E\left(\left.\boldsymbol\xi_{it}f_{tk}\right| \boldsymbol \Gamma\right)\E\left(\left.\boldsymbol\xi_{is}f_{sk^{\prime}}\right| \boldsymbol \Gamma\right)^{\top}\right].
\end{eqnarray*}

Note that for the case $\vartheta=1$ the cross-sectional CLT term and the strong mixing CLT term are of the same order. However, we have by Assumption 3.6 (i) that $\{ \mathbf{u}_t\}_{t\leq T}$ is independent of $\{\mathbf{Z}_i,\boldsymbol{\xi}_t,\mathbf{f}_t, \boldsymbol{\gamma}_{i} \}_{i\leq N; t\leq T}$. Therefore the left-hand side terms of (\ref{eqbb2}) and (\ref{eqbb3}) are independent.

Combining (i)-(iii) with (\ref{eqA3}), the proof of the Theorem is done.
}

\subsection{Proof of Theorem \ref{theo_bootstrap}}\label{appendix_B2}
We denote every statistics object in the bootstrap case conditioning on the original sample with a $*$ from now on. We need to show that the bootstrap version of the estimator $v^\top(\widehat{\boldsymbol{\beta}}_{b}^{*}-\widehat{\boldsymbol{\beta}})$ has the same asymptotic normal distribution.
Let $\widetilde{\boldsymbol{\beta}}_{b}^{*}$ denote the bootstrap counterpart of $\widetilde{\boldsymbol{\beta}}_{b}$  as defined in Lemma A.7. We denote $\mathcal{O}_{p^*}(1)$ as the probability limit corresponding to the probability measure with respect to the original sample. Similarly to the derivation in Theorem \ref{theo1}, we have the following expansion with the bootstrap measure conditioning on the original sample,
\begin{align*}
 & v^{\top}(\widehat{\boldsymbol{\beta}}_{b}^{*}-\widehat{\boldsymbol{\beta}})=v^{\top}(\widetilde{\boldsymbol{\beta}}_{b}^{*}-\widehat{\boldsymbol{\beta}})+{\scriptstyle\mathcal{O}}_{p^*}\left(\frac{1}{\sqrt{NT}}\right)+{\scriptstyle\mathcal{O}}_{p^*}\left(\frac{\sqrt{\nu_N}}{\sqrt{N}}\right),
\end{align*}
and
\begin{align*}
v^{\top}(\widetilde{\boldsymbol{\beta}}_b^*-
\widehat{\boldsymbol{\beta}}) & =v^{\top}S_{X-\widetilde{X}}^{*-1}\left(S_{X-\widetilde{X},(G-\widetilde{G})f}^{*}+S_{X-\widetilde{X},V-\widetilde{V}}^{*}\right)\\
 & =v^{\top}S_{X-\widetilde{X}}^{*-1}S_{X-\widetilde{X},V}^{*}+\Co_{p^{*}}(1/\sqrt{NT}).
\end{align*}

Let $n=NT$ and $(i,t)\to k$. Now we prove the bootstrap quantile
is valid. We cite lemma A.7 which is from Theorem 1 in \cite{sunklodas1984rate}.

Let $X_{k}$ be the summand within $v^{\top}S_{X-\widetilde{X}}^{*-1}S_{X-\widetilde{X},V}^{*}=\sum_{t}
v^{\top}\{S_{X-\widetilde{X}}^{-1}[\frac{1}{NT}
\boldsymbol\xi_{t}^{^{*}\top}\mathbf{u}_{t}^{*} \\ +\frac{1}{NT}\E^{*}\left(
\boldsymbol\xi_{t}^{^{*}\top}\boldsymbol\Gamma^{*}\mathbf{f}_{t}^{*}
\boldsymbol \Gamma^{*}\right)]\}+\Co_{p^{*}}(1/\sqrt{NT})$.

Consider Assumption \ref{assum:covariance} and let $B_{n}^{*}=S_{n}^{1/2}$. Define $W_v=v^{\top}W$ with $W$ being a $Q$-dimensional standard normal random variable. Let $P^*$ be any measure conditional on $\{X_{it},Y_{it}\}$. By Lemma A.7,
\[
\sup_{x}\left|\P^{*}\left(V_{\boldsymbol{\beta},v,n}^{1/2}W_v\leq x\right)-\P^{*}\left(\sqrt{NT}v^{\top}\left(\boldsymbol{\widehat{\boldsymbol{\beta}}}_{b}^{*}-\boldsymbol{\widehat{\boldsymbol{\beta}}}\right)\leq x\right)\right|=\CO_{p^*}\left(\E{_{n}}\left|X_{k}\right|^{p}\left(\log B_{n}^{*}\right)^{p-1}B_{n}^{*2-p}\right)
\]

This follows from the Berry-Esseen theorem, as long as the following
holds, with probability approaching $1$ and uniformly over $\P\in\mathcal{P},$
exists a positive constant $c$, we have the following statement,
\[
\frac{1}{n}\sum_{k=1}^n\left|X_{k}\right|^{p}\leq c\E\left|X_{k}\right|^{p},\quad B_{n}^{*}\leq cB_{n}.
\]
(This statement follows the law of large numbers and is maintained by our Assumption \ref{asum5}) Take $p=2+\delta$ , then $\CO_p\left(\E_{n}\left|X_{k}\right|^{p}\left(\log B_{n}^{*}\right)^{p-1}B_{n}^{*2-p}\right)=\CO_p((\log n)^{1+\delta}n^{-\delta/2})=\CO_p(1)$ followed by Assumption \ref{asum5}.

Next, we shall prove that,
\begin{eqnarray}\label{eqBoot1}
\sup_{x}\left|\P^{*}\left(V_{\beta,v,n}^{1/2}W_{v}\leq x\right)-\P\left(V_{\beta,v}^{1/2}W_{v}\leq x\right)\right|=\Co_{p}(1),
\end{eqnarray}

Let $f^{*}(.)$ be a density function conditional on the original sample.
Take a Taylor expansion of $\P^{*}\left(W_{v}\leq V_{\beta,v,n}^{-1/2}x\right)$ we have
\[
\P^{*}\left(W_{\mathbf{v}}\leq V_{\beta,v,n}^{-1/2}x\right)=\P^{*}\left(W_{v}\leq V_{\beta,v}^{-1/2}x\right)+f^{*}\left(W_{v}\leq V_{\beta,v}^{-1/2}\widetilde{x}\right)(V_{\beta,v}^{-1/2}x-V_{\beta,v,n}^{-1/2}x),
\]
where $V_{\beta,v}^{-1/2}\widetilde{x}$ is a mid point between $V_{\beta,v}^{-1/2}x$
and $V_{\beta,v,n}^{-1/2}x$. By Assumption \ref{assum:covariance} we have $(V_{\beta,v}^{-1/2}-V_{\beta,v,n}^{-1/2})=\Co(1)$.
Then statement (\ref{eqBoot1}) is true.

The result (\ref{eqBoot2}) for a fixed sequence follows the fact that
\[
\P^{*}\left(W_{v}\leq V_{\beta,v}^{-1/2}x\right)=\P\left(W_{v}\leq V_{\beta,v}^{-1/2}x\right)
\]
and
\[
(V_{\beta,v}^{-1/2}x-V_{\beta,v,n}^{-1/2}x)=\Co(1).
\]

Thus, we can show that uniformly for $\P\in\mathcal{P}$,
\[
\sup_{x}\left|\P\left(W_{v}\leq V_{\beta,v}^{-1/2}x\right)-\P\left(V_{\beta,v}^{1/2}W_{v}\leq x\right)\right|=\Co(1).
\]

Then similarly from the steps of the proof of Proposition H.1 (iii) in \citet{FernndezVal2022DynamicHD}, we can show that uniformly for $\P\in\mathcal{P}$,
\begin{eqnarray}\label{eqBoot2}
\P\left(v^{\top}(\widehat{\boldsymbol{\beta}}-\boldsymbol{\beta})\leq q_{\alpha,\mathbf{v}}\right)\rightarrow \P^{*}\left(v^{\top}\left(\widehat{\boldsymbol{\beta}}_{b}^{*}-
\widehat{\boldsymbol{\beta}}\right)\leq q_{\alpha,\mathbf{v}}\right)=1-\alpha.
\end{eqnarray}
 We list the steps as follows: consider sequences of probability laws
$\left\{ \P_{T}:T\geq1\right\} \subset\mathcal{P}$ with sub-sequences.

Therefore for the event: $E:=\left\{ v^{\top}\boldsymbol{\beta}\in CI_{\alpha}\left(v^{\top}\boldsymbol{\beta}\right)\right\} $,
we have
\[
\P\left(v^{\top}(\widehat{\boldsymbol{\beta}}-\boldsymbol{\beta})\leq q_{\alpha,\mathbf{v}}\right)\rightarrow \P^{*}\left(v^{\top}\left(\widehat{\boldsymbol{\beta}}_{b}^{*}-
\widehat{\boldsymbol{\beta}}\right)\leq q_{\alpha,v}\right)=1-\alpha
\]

Now consider a DGP sequence $\left\{ \P_{n}:n\geq1\right\} \subset\mathcal{P}$
and its subsequence $\left\{ \P_{n_{k}}:n\geq1\right\} $ so that
\[
\limsup_n\sup_{\P\in\mathcal{P}}|\P(E)-(1-\alpha)|=\limsup_{n}\left|\P_{n}(E)-(1-\alpha)\right|=\lim_{k}\left|\P_{n_{k}}(E)-(1-\alpha)\right|.
\]

Note that $\left\{ \P_{n_{k}}:k\geq1\right\} \subset\mathcal{P}$ so the probability measure $\P_{n_{k}}$ satisfies the conditions of
our previous derivation by changing $n$ to $n_{k}$. It implies $\lim_{k}\P_{n_{k}}(E)\rightarrow1-\alpha$.
Hence
\[
\limsup_{T}\sup_{P\in\mathcal{P}}|\P(E)-(1-\alpha)|=0
\]

\section{Estimation of Latent Factors and Loadings}\label{appendix_C}
In this subsection, we show how to estimate the factors and the loadings once we have available an estimate for the slope parameters. Note however that for the estimation of $\boldsymbol{\beta}$, as it has been already pointed out, it is not necessary to estimate previously $\mathbf{f}_t$ or $\boldsymbol{\boldsymbol \lambda}_i$.  To estimate the latent factors and corresponding factor loadings we propose to use the projected PCA approach proposed in \cite{Fan-Liao-Wang2016} in a slightly different context. More precisely, owing to potential correlations between the unobservable effects and the regressors, we treat the matrix of common factors $\mathbf F=(\mathbf f_1,\ldots,\mathbf{f}_T)^{\top}$ as fixed-effects parameters to be estimated.
The latent factors and corresponding factor loadings can be estimated from the regression residuals, $\mathbf{\widetilde{y}}_{t}=\mathbf y_{t}-\mathbf X_{t}\boldsymbol{\boldsymbol{\widehat{\boldsymbol{\beta}}}}$,
and let $\mathbf{\widetilde{Y}}=(\mathbf{\widetilde{y}}_{1},\ldots,\mathbf{\widetilde{y}}_{T})$. Now $\mathbf F$ and $\mathbf{G}(Z)$ can be recovered from the projected data $\mathbf{P}_{\Phi}(\mathbf{Z})\mathbf{\widetilde{Y}}$ under the following identification restrictions: (i) $T^{-1}\mathbf F^{\top}\mathbf F=I_K$ and (ii) $\mathbf{G}(\mathbf{Z})^{\top}\mathbf{G}(\mathbf{Z})$ is a $K\times K$ diagonal matrix with distinct entries. Then, $\sqrt{T}\mathbf{\widehat{F}}$ can be estimated by the eigenvectors associated with the largest $K$ eigenvalues of the matrix $T^{-1}\mathbf{\widetilde{Y}}^{\top}\mathbf P_{\Phi}(\mathbf{Z})\mathbf{\widetilde{Y}}$.

Once we estimated $\mathbf{\widehat{F}}$, it is possible to obtain an estimator for the sieve coefficients using a least squares procedure that leads to
\begin{align}\label{eq14} \mathbf{\widehat{B}}=(\mathbf {\widehat{b}}_{1},\ldots,\mathbf{\widehat{b}}_{K})=\frac{1}{T}\left[\Phi(\mathbf{Z})^{\top}\Phi(\mathbf{Z})\right]^{-1}\Phi(\mathbf{Z})^{\top}\mathbf{\widetilde{Y}}\mathbf{\widehat{F}},
\end{align}
and replacing (\ref{eq14}) into (\ref{eq7}), we can construct an estimator for the nonparametric functions $g_k(\cdot)$ of the form
\begin{align}\label{eq15}
	\widehat{g}_{k}(z)=\phi(z)^{\top}\mathbf{\widehat{b}}_{k},\quad k=1,\ldots,K.
\end{align}

Let $\boldsymbol{\boldsymbol \Lambda}$ be the $N\times K$ matrix of factor loadings, then we can estimate $\boldsymbol{\widehat{\boldsymbol \Lambda}}=\mathbf{\widetilde{Y}}\mathbf{\widehat{F}}/T$ (see \citet{Fan-Liao-Wang2016} for further details). Therefore, the part of the factor loadings in (\ref{eq3}) that can be explained by $\mathbf{Z}$ can be estimated by $\mathbf{\widehat{G}}(\mathbf{Z})=\frac{1}{T}\mathbf P_{\Phi}(\mathbf{Z})\mathbf{\widetilde{Y}}\mathbf{\widehat{F}}$, whereas the idiosyncratic part can be calculated as $\boldsymbol{\widehat{\boldsymbol \Gamma}}=\boldsymbol{\widehat{\boldsymbol \Lambda}}-\mathbf{\widehat{G}}(\mathbf{Z})=\frac{1}{T}[\mathbf I_{N}-\mathbf P_{\Phi}(\mathbf{Z})]\mathbf{\widetilde{Y}}\mathbf{\widehat{F}}$.

The consistency result is established in the following proposition. The proof is adapted from the Theorem 4.1 of \citet{Fan-Liao-Wang2016} to our case of time-varying covariates. We have to impose the following additional assumptions.
\begin{assumption}[Identification]\label{asum22}
Almost surely, $T^{-1}\mathbf F^{\top}\mathbf F=\mathbf I_K$ and $\mathbf G(\mathbf{Z})^{\top}\mathbf G(\mathbf{Z})$ is a $K\times K$ diagonal matrix with distinct entries.
\end{assumption}

The condition detailed above is commonly used in the estimation of factor models and corresponds to condition PC1 of \citet{Bai-Ng2013} in the case of unprojected data. In particular, it enables to identify separately the factors $\mathbf F$ and the part of the loadings explained by the covariates, $\mathbf G(\mathbf{Z})$, from their product $\mathbf G(\mathbf{Z})\mathbf F$.

\begin{assumption}[Genuine projection]\label{asum111}
There are two positive constants, $c_{\min}$ and $c_{\max}$ such that, with probability approaching one (as $N \rightarrow \infty$),
\begin{eqnarray*}
c_{\min} < \lambda_{min}\left[N^{-1}\mathbf G(\mathbf{Z})^{\top}\mathbf G(\mathbf{Z})\right] <
\lambda_{max}\left[N^{-1}\mathbf G(\mathbf{Z})^{\top} \mathbf G(\mathbf{Z})\right] <c_{\max}.
\end{eqnarray*}
\end{assumption}

\begin{assumption}[Data generating process]\label{asum_factor}\hfill
\begin{description}
\item[(i)] Exponential mixing rate:
there exists $r_1,C_1>0$ such that for all $T>0$,
\begin{align*}
    \alpha(T)<\exp\left(-C_1T\right).
\end{align*}
\item[(ii)] Exponential tail: there exist $r_2,r_3>0$ satisfying $r_1^{-1}+r_2^{-1}+r_3^{-1}>1$ and $b_1,b_2>0$, such that for any $x>0$, $i\leq N$ and $k\leq K$,
    \begin{align*}
    \P\left(|u_{it}|>z\right)\leq\exp\left\{-\left(z/b_1\right)^{r_2}\right\},\\
    \P\left(|f_{kt}|>z\right)\leq\exp\left\{-\left(z/b_2\right)^{r_3}\right\}.
\end{align*}
\end{description}
\end{assumption}

Assumption \ref{asum111} is the key condition of the projected PCA and it is similar to the pervasive condition on the factor loadings assumed in \citet{Stock-Watson2002}. In particular, it requires that the $\mathbf{Z}$'s have non-vanishing explaining power on the systematic part of the loadings, $\mathbf G(\cdot)$, so $\mathbf G(\mathbf{Z})^{\top}\mathbf G(\mathbf{Z})$ has spiked eigenvalues and rules out the case when $\mathbf{Z}$ is completely unassociated with $\boldsymbol \Lambda$ (i.e., when $\mathbf{Z}$ is pure noise).
{
\begin{prop}\label{theo2}
Under Assumptions \ref{asum0}-\ref{asum66}, \ref{asum22}-\ref{asum_factor}, we have that, as both $N$ and $T$ tend to infinity,
\begin{align*}
\frac{1}{T}\|\widehat{\mathbf F}-\mathbf F\|_{F}^{2}&=\mathcal{O}_{p}\left(\frac{1}{N}+\frac{1}{J_N^{\kappa}}\right),\\
\frac{1}{N}\|\widehat{\mathbf G}(\mathbf{Z})-\mathbf{G}(\mathbf{Z})\|_{F}^{2}&=\mathcal{O}_{p}
\left(\frac{J_N}{N^2}+\frac{J_N}{NT}+\frac{J_N}{J_N^{\kappa}}+\frac{J_N\nu_N}{N}\right),\\
\max_{k=1,\ldots,K}\sup_{z\in\mathcal{Z}}\left|\widehat{g}_{k}(z)-g_{k}(z)\right|&
=\mathcal{O}_{p}\left(\frac{J_N}{N}+\frac{J_N}{\sqrt{NT}}+\frac{J_N}{J_N^{\kappa/2}}+
J_N\sqrt{\frac{\nu_N}{N}}\right).
\end{align*}
\end{prop}}

As it can be seen in Theorem \ref{theo2}, the convergence rates of the estimators of the interactive fixed effects components are not affected by $\boldsymbol{\widehat{\boldsymbol{\beta}}}$. That is, they are identical to the pure factor model case of \citet{Fan-Liao-Wang2016}. This follows from the convergence rate of $\boldsymbol{\widehat{\boldsymbol{\beta}}}$ from our Theorem \ref{theo1}. Furthermore, the above results have been obtained assuming $T$ is large, but that is not required and similar results can be obtained for small $T$.

\section{Additional Simulation Results}\label{appendix_D}
As a robustness check, we consider the case of $t$-distributed error terms with $5$ degrees of freedom. The results reported in Table \ref{table:rmse_t} show that the estimation accuracy can be effectively improved with growing sample size even in the case of error with wider tails.
\setcounter{table}{0}
\renewcommand{\thetable}{A\arabic{table}}
\begin{table}[tbp]
\spacingset{1.2} 
\centering
\begin{tabular}{lrr|rrr|rrr}
	\hline \hline
  \multicolumn{2}{r}{} & \multicolumn{3}{c}{$\boldsymbol{\beta}_1$} & \multicolumn{3}{c}{$\boldsymbol{\beta}_2$} \\
  $\nu_N$ & $N$ & $T$ & P-IFE & PC-IFE & bc-PC-IFE & P-IFE & PC-IFE & bc-PC-IFE \\
  \hline
 $\nu_N=\mathcal{O}(1)$ & 50 & 10 & 0.0855 & 0.0703 & 0.0703 & 0.0876 & 0.0733 & 0.0733 \\ 
  & 100 & 10 & 0.0560 & 0.0539 & 0.0541 & 0.0597 & 0.0515 & 0.0514 \\ 
  & 50 & 50 & 0.0530 & 0.0302 & 0.0303 & 0.0554 & 0.0290 & 0.0291 \\ 
  & 100 & 50 & 0.0363 & 0.0243 & 0.0244 & 0.0345 & 0.0236 & 0.0238 \\ 
  & 200 & 100 & 0.0212 & 0.0147 & 0.0149 & 0.0205 & 0.0140 & 0.0140 \\ 
  & 500 & 100 & 0.0133 & 0.0125 & 0.0129 & 0.0132 & 0.0123 & 0.0125 \\ 
  \hline
  $\nu_N=0$ & 50 & 10 & 0.0551 & 0.0755 & 0.0754 & 0.0521 & 0.0728 & 0.0728 \\ 
  & 100 & 10 & 0.0380 & 0.0525 & 0.0524 & 0.0357 & 0.0503 & 0.0503 \\ 
  & 50 & 50 & 0.0235 & 0.0309 & 0.0310 & 0.0246 & 0.0313 & 0.0314 \\ 
  & 100 & 50 & 0.0169 & 0.0240 & 0.0242 & 0.0159 & 0.0229 & 0.0229 \\ 
  & 200 & 100 & 0.0078 & 0.0156 & 0.0159 & 0.0085 & 0.0155 & 0.0155 \\ 
  & 500 & 100 & 0.0051 & 0.0154 & 0.0156 & 0.0051 & 0.0148 & 0.0150 \\ 
  \hline
  $\nu_N=\mathcal{O}(T^{-1})$ & 50 & 10 & 0.0577 & 0.0698 & 0.0699 & 0.0601 & 0.0693 & 0.0693 \\ 
  & 100 & 10 & 0.0399 & 0.0490 & 0.0490 & 0.0419 & 0.0449 & 0.0448 \\ 
  & 50 & 50 & 0.0252 & 0.0232 & 0.0236 & 0.0262 & 0.0245 & 0.0248 \\ 
  & 100 & 50 & 0.0175 & 0.0179 & 0.0181 & 0.0169 & 0.0175 & 0.0177 \\ 
  & 200 & 100 & 0.0083 & 0.0092 & 0.0097 & 0.0085 & 0.0090 & 0.0093 \\ 
  & 500 & 100 & 0.0052 & 0.0069 & 0.0073 & 0.0051 & 0.0068 & 0.0071 \\ 
 \hline \hline
\end{tabular}
\caption{$RMSE$ of the projected IFE estimator, the PC-based IFE, and the bias-corrected PC-based IFE estimator under $t_{5}$-distributed error terms.}
\label{table:rmse_t}
\end{table}

Additionally, we consider another data-generating process with a larger number of factors, $K=10$. For the systematic part of the factor loadings, $g_{k}$, we consider: $g_1(z)=\sin(2z_1)^3+z_2^2$, $g_2(z)=-z_1^2+\cos(2z_2)$, $g_3(z)=\tan(z_2^3)-3z_1$, $g_4(z)=2z_1^4$, $g_5(z)=-\sin(3z_2)^2+2z_1$, $g_6(z)=z_2^4$, $g_7(z)=2\cos(z_1)^3+z_2$, $g_8(z)=-z_1^3+2z_2^2$, $g_9(z)=\sin(z_1^3)-2\cos(z_1)$, $g_{10}(z)=z_2-z_1$. The results in Table \ref{table:rmse_factor} for i.i.d. data and in Table \ref{table:rmse_factor_AR} for serially correlated error terms show that our estimator still performs best in all the settings for $N$ and $T$ in settings with $\nu_N=0$ and $\nu_N=\mathcal{O}(T^{-1})$. Additionally, the P-IFE outperforms the PC-IFE in the shot panel setting with $T=10$.

\begin{table}[tbp]
\spacingset{1.2} 
\centering
\begin{tabular}{lrr|rrr|rrr}
	\hline \hline
  \multicolumn{2}{r}{} & \multicolumn{3}{c}{$\boldsymbol{\beta}_1$} & \multicolumn{3}{c}{$\boldsymbol{\beta}_2$} \\
  $\nu_N$ & $N$ & $T$ & P-IFE & PC-IFE & bc-PC-IFE & P-IFE & PC-IFE & bc-PC-IFE \\
  \hline
 $\nu_N=\mathcal{O}(1)$ & 50 & 10 & 0.0863 & 0.1104 & 0.1105 & 0.0824 & 0.1138 & 0.1136 \\ 
  & 100 & 10 & 0.0563 & 0.0924 & 0.0925 & 0.0534 & 0.0925 & 0.0924 \\ 
  & 50 & 50 & 0.0508 & 0.0351 & 0.0351 & 0.0492 & 0.0348 & 0.0349 \\ 
  & 100 & 50 & 0.0344 & 0.0264 & 0.0265 & 0.0326 & 0.0268 & 0.0268 \\ 
  & 200 & 100 & 0.0193 & 0.0133 & 0.0137 & 0.0201 & 0.0148 & 0.0149 \\ 
  & 500 & 100 & 0.0116 & 0.0131 & 0.0134 & 0.0117 & 0.0117 & 0.0117 \\ 
  \hline
  $\nu_N=0$ & 50 & 10 & 0.0363 & 0.0918 & 0.0917 & 0.0363 & 0.0895 & 0.0897 \\ 
  & 100 & 10 & 0.0228 & 0.0797 & 0.0797 & 0.0229 & 0.0765 & 0.0765 \\ 
  & 50 & 50 & 0.0155 & 0.0203 & 0.0204 & 0.0173 & 0.0215 & 0.0217 \\ 
  & 100 & 50 & 0.0111 & 0.0151 & 0.0155 & 0.0106 & 0.0151 & 0.0151 \\ 
  & 200 & 100 & 0.0053 & 0.0082 & 0.0088 & 0.0050 & 0.0080 & 0.0082 \\ 
  & 500 & 100 & 0.0033 & 0.0065 & 0.0073 & 0.0030 & 0.0063 & 0.0066 \\ 
  \hline
  $\nu_N=\mathcal{O}(T^{-1})$ & 50 & 10 & 0.0412 & 0.0573 & 0.0575 & 0.0438 & 0.0537 & 0.0538 \\ 
  & 100 & 10 & 0.0284 & 0.0423 & 0.0425 & 0.0277 & 0.0407 & 0.0408 \\ 
  & 50 & 50 & 0.0165 & 0.0173 & 0.0176 & 0.0166 & 0.0170 & 0.0172 \\ 
  & 100 & 50 & 0.0111 & 0.0131 & 0.0134 & 0.0112 & 0.0133 & 0.0136 \\ 
  & 200 & 100 & 0.0056 & 0.0066 & 0.0071 & 0.0054 & 0.0074 & 0.0077 \\ 
  & 500 & 100 & 0.0031 & 0.0055 & 0.0060 & 0.0035 & 0.0053 & 0.0055 \\ 
 \hline \hline
\end{tabular}
\caption{$RMSE$ of the projected IFE estimator, the PC-based IFE, and the bias-corrected PC-based IFE estimator in the setting with $K=10$ under i.i.d. error terms.}
\label{table:rmse_factor}
\end{table}

\begin{table}[tbp]
\spacingset{1.2} 
\centering
\begin{tabular}{lrr|rrr|rrr}
	\hline \hline
  \multicolumn{2}{r}{} & \multicolumn{3}{c}{$\boldsymbol{\beta}_1$} & \multicolumn{3}{c}{$\boldsymbol{\beta}_2$} \\
  $\nu_N$ & $N$ & $T$ & P-IFE & PC-IFE & bc-PC-IFE & P-IFE & PC-IFE & bc-PC-IFE \\
  \hline
 $\nu_N=\mathcal{O}(1)$ & 50 & 10 & 0.0845 & 0.1115 & 0.1113 & 0.0848 & 0.1032 & 0.1032 \\ 
  & 100 & 10 & 0.0557 & 0.0932 & 0.0931 & 0.0538 & 0.0952 & 0.0950 \\ 
  & 50 & 50 & 0.0486 & 0.0370 & 0.0372 & 0.0484 & 0.0346 & 0.0345 \\ 
  & 100 & 50 & 0.0326 & 0.0263 & 0.0264 & 0.0329 & 0.0269 & 0.0270 \\ 
  & 200 & 100 & 0.0192 & 0.0138 & 0.0140 & 0.0192 & 0.0142 & 0.0144 \\ 
  & 500 & 100 & 0.0124 & 0.0120 & 0.0124 & 0.0125 & 0.0130 & 0.0130 \\ 
  \hline
  $\nu_N=0$ & 50 & 10 & 0.0349 & 0.0879 & 0.0881 & 0.0351 & 0.0905 & 0.0906 \\ 
  & 100 & 10 & 0.0224 & 0.0797 & 0.0799 & 0.0230 & 0.0791 & 0.0792 \\ 
  & 50 & 50 & 0.0161 & 0.0216 & 0.0216 & 0.0166 & 0.0219 & 0.0221 \\ 
  & 100 & 50 & 0.0100 & 0.0147 & 0.0151 & 0.0105 & 0.0149 & 0.0150 \\ 
  & 200 & 100 & 0.0053 & 0.0083 & 0.0089 & 0.0053 & 0.0082 & 0.0084 \\ 
  & 500 & 100 & 0.0032 & 0.0063 & 0.0072 & 0.0033 & 0.0065 & 0.0069 \\ 
  \hline
  $\nu_N=\mathcal{O}(T^{-1})$ & 50 & 10 & 0.0419 & 0.0555 & 0.0555 & 0.0424 & 0.0531 & 0.0530 \\ 
  & 100 & 10 & 0.0287 & 0.0425 & 0.0426 & 0.0277 & 0.0396 & 0.0396 \\ 
  & 50 & 50 & 0.0163 & 0.0173 & 0.0176 & 0.0164 & 0.0166 & 0.0168 \\ 
  & 100 & 50 & 0.0112 & 0.0131 & 0.0134 & 0.0111 & 0.0135 & 0.0137 \\ 
  & 200 & 100 & 0.0051 & 0.0071 & 0.0078 & 0.0053 & 0.0069 & 0.0073 \\ 
  & 500 & 100 & 0.0033 & 0.0057 & 0.0062 & 0.0033 & 0.0056 & 0.0059 \\ 
 \hline \hline
\end{tabular}
\caption{$RMSE$ of the projected IFE estimator, the PC-based IFE, and the bias-corrected PC-based IFE estimator in the setting with $K=10$ and under autocorrelated Gaussian error terms (MA($\infty$) with algebraic decay parameter $5$.}
\label{table:rmse_factor_AR}
\end{table}

Finally, we also observe good coverage results for our cross-sectional bootstrap procedure in the case of $t$-distributed error terms in Table \ref{table:coverage_t} and in the case of auto-correlated error terms in Table \ref{table:coverage_AR}. Effectively, the empirical coverage approaches the nominal coverage with increasing sample size. We therefore shown the robustness of the cross-sectional bootstrap procedure towards serially correlated errors and error terms with heavier tail.

\begin{table}[tbp]
\centering
\begin{tabular}{lrr|rrr|rrr}
	\hline \hline
  \multicolumn{2}{r}{} & \multicolumn{3}{c}{$\boldsymbol{\beta}_1$} & \multicolumn{3}{c}{$\boldsymbol{\beta}_2$} \\
  $\nu_N$ & $N$ & $T$ & $90\%$ & $95\%$ & $99\%$ & $90\%$ & $95\%$ & $99\%$ \\
  \hline
  $\nu_N=\mathcal{O}(1)$ & 50 & 10 & 0.844 & 0.916 & 0.976 & 0.886 & 0.932 & 0.968 \\ 
  & 100 & 10 & 0.862 & 0.932 & 0.986 & 0.902 & 0.954 & 0.988 \\ 
  & 50 & 50 & 0.832 & 0.898 & 0.974 & 0.864 & 0.924 & 0.978 \\ 
  & 100 & 50 & 0.850 & 0.922 & 0.978 & 0.858 & 0.914 & 0.980 \\ 
  & 200 & 100 & 0.912 & 0.952 & 0.990 & 0.870 & 0.934 & 0.980 \\ 
  & 500 & 100 & 0.912 & 0.952 & 0.992 & 0.868 & 0.928 & 0.986 \\ 
  \hline
  $\nu_N=0$ & 50 & 10 & 0.866 & 0.932 & 0.966 & 0.874 & 0.920 & 0.978 \\ 
  & 100 & 10 & 0.876 & 0.940 & 0.970 & 0.866 & 0.932 & 0.978 \\ 
  & 50 & 50 & 0.882 & 0.920 & 0.976 & 0.882 & 0.934 & 0.972 \\ 
  & 100 & 50 & 0.900 & 0.936 & 0.980 & 0.888 & 0.940 & 0.984 \\ 
  & 200 & 100 & 0.880 & 0.924 & 0.986 & 0.870 & 0.924 & 0.978 \\ 
  & 500 & 100 & 0.898 & 0.956 & 0.990 & 0.884 & 0.940 & 0.996 \\ 
  \hline
  $\nu_N=\mathcal{O}(T^{-1})$ & 50 & 10 & 0.868 & 0.926 & 0.978 & 0.874 & 0.932 & 0.978 \\ 
  & 100 & 10 & 0.870 & 0.926 & 0.978 & 0.878 & 0.936 & 0.984 \\ 
  & 50 & 50 & 0.870 & 0.920 & 0.972 & 0.896 & 0.934 & 0.980 \\ 
  & 100 & 50 & 0.868 & 0.934 & 0.978 & 0.902 & 0.940 & 0.988 \\ 
  & 200 & 100 & 0.866 & 0.914 & 0.982 & 0.898 & 0.934 & 0.978 \\ 
  & 500 & 100 & 0.902 & 0.944 & 0.998 & 0.908 & 0.958 & 0.996 \\ 
 \hline \hline
\end{tabular}
\caption{Empirical coverage probabilities of the cross-sectional bootstrap confidence intervals under $t_{5}$-distributed errors. Results based on $B=1000$ bootstrap iterations.}
\label{table:coverage_t}
\end{table}

\begin{table}[tbp]
\centering
\begin{tabular}{lrr|rrr|rrr}
	\hline \hline
  \multicolumn{2}{r}{} & \multicolumn{3}{c}{$\boldsymbol{\beta}_1$} & \multicolumn{3}{c}{$\boldsymbol{\beta}_2$} \\
  $\nu_N$ & $N$ & $T$ & $90\%$ & $95\%$ & $99\%$ & $90\%$ & $95\%$ & $99\%$ \\
  \hline
  $\nu_N=\mathcal{O}(1)$ & 50 & 10 & 0.892 & 0.928 & 0.964 & 0.854 & 0.912 & 0.970 \\ 
  & 100 & 10 & 0.872 & 0.924 & 0.984 & 0.860 & 0.920 & 0.974 \\ 
  & 50 & 50 & 0.826 & 0.906 & 0.970 & 0.850 & 0.926 & 0.976 \\ 
  & 100 & 50 & 0.866 & 0.924 & 0.982 & 0.888 & 0.938 & 0.986 \\ 
  & 200 & 100 & 0.876 & 0.934 & 0.986 & 0.902 & 0.948 & 0.986 \\ 
  & 500 & 100 & 0.894 & 0.944 & 0.986 & 0.880 & 0.938 & 0.984 \\ 
  \hline
  $\nu_N=0$ & 50 & 10 & 0.832 & 0.916 & 0.962 & 0.864 & 0.920 & 0.976 \\ 
  & 100 & 10 & 0.882 & 0.932 & 0.984 & 0.856 & 0.918 & 0.986 \\ 
  & 50 & 50 & 0.872 & 0.934 & 0.988 & 0.884 & 0.920 & 0.964 \\ 
  & 100 & 50 & 0.862 & 0.928 & 0.984 & 0.866 & 0.914 & 0.982 \\ 
  & 200 & 100 & 0.884 & 0.938 & 0.990 & 0.900 & 0.946 & 0.984 \\ 
  & 500 & 100 & 0.904 & 0.952 & 0.992 & 0.882 & 0.930 & 0.984 \\
  \hline
  $\nu_N=\mathcal{O}(T^{-1})$ & 50 & 10 & 0.846 & 0.906 & 0.978 & 0.900 & 0.948 & 0.986 \\ 
  & 100 & 10 & 0.862 & 0.920 & 0.982 & 0.878 & 0.930 & 0.974 \\ 
  & 50 & 50 & 0.842 & 0.906 & 0.982 & 0.888 & 0.936 & 0.988 \\ 
  & 100 & 50 & 0.884 & 0.934 & 0.980 & 0.870 & 0.930 & 0.982 \\ 
  & 200 & 100 & 0.884 & 0.924 & 0.978 & 0.904 & 0.958 & 0.994 \\ 
  & 500 & 100 & 0.876 & 0.942 & 0.980 & 0.900 & 0.952 & 0.990 \\ 
 \hline \hline
\end{tabular}
\caption{Empirical coverage probabilities of the cross-sectional bootstrap confidence intervals autocorrelated Gaussian error terms (MA($\infty$) with algebraic decay parameter $5$. Results based on $B=1000$ bootstrap iterations.}
\label{table:coverage_AR}
\end{table}

\end{document}